\documentclass[preprint,12pt]{elsarticle}

\usepackage{graphicx}
\usepackage{amssymb}

\usepackage{lineno}

\journal{arXiv}

\usepackage[section]{placeins}

\usepackage{framed}
\usepackage{multicol}
\usepackage{multirow}
\usepackage{makecell}
\usepackage{nomencl}
\usepackage{bm}
\usepackage{amsmath}
\usepackage[english]{babel}

\usepackage{hyperref} 
\hypersetup{
    colorlinks=false,
    linkcolor=black,
    filecolor=magenta,      
    urlcolor=cyan,
}
\AtBeginDocument{\hypersetup{pdfborder={0 0 1}}}

\begin{document}
\begin{frontmatter}

\title{Topographic uncertainty quantification for flow-like landslide models via stochastic simulations}



\author[Aachen]{Hu Zhao}
\ead{zhao@aices.rwth-aachen.de}

\author[Aachen]{Julia Kowalski}

\address[Aachen]{RWTH Aachen University, Templergraben 55, 52062 Aachen, Germany}

\begin{abstract}
Topography representing digital elevation models (DEMs) are essential inputs for computational models capable of simulating the run-out of flow-like landslides. Yet, DEMs are often subject to error, a fact that is mostly overlooked in landslide modeling. We address this research gap and investigate the impact of topographic uncertainty on landslide-run-out models. In particular, we will describe two different approaches to account for DEM uncertainty, namely unconditional and conditional stochastic simulation methods. We investigate and discuss their feasibility, as well as whether DEM uncertainty represented by stochastic simulations critically affects landslide run-out simulations. Based upon a historic flow-like landslide event in Hong Kong, we present a series of computational scenarios to compare both methods using our modular Python-based workflow. Our results show that DEM uncertainty can significantly affect simulation-based landslide run-out analyses, depending on how well the underlying flow path is captured by the DEM, as well as further topographic characteristics and the DEM error's variability. We further find that in the absence of systematic bias in the DEM, a performant root mean square error based unconditional stochastic simulation yields similar results than a computationally intensive conditional stochastic simulation that takes actual DEM error values at reference locations into account. In all other cases the unconditional stochastic simulation overestimates the variability of the DEM error, which leads to an increase  of the potential hazard area as well as extreme values of dynamic flow properties.
\end{abstract}


\end{frontmatter}


\section{Introduction}
\label{S:1}
Landslides are natural hazards that occur frequently all around the world causing casualties, economic devastation, and environmental destruction. Most often, they are naturally driven, e.g. by means of long-lasting and/or intensive precipitation events, or induced by earthquakes. Yet, landslides might also be triggered or its susceptibility increased as a result of human activities, e.g. deforestation and construction. According to the United Nations Office for Disaster Risk Reduction and the Center for Research on the Epidemiology of Disasters, 378 recorded landslides from 1998 to 2017 affected 4.8 million people and caused 18414 deaths as well as several billion US dollars of economic losses \citep{Report2018}. \citet{Froude2018} reported that in total 55997 people were killed during 4862 fatal non-seismic landslide events from January 2004 to December 2016. Still, it has to be assumed that the damage potential of landslides is underestimated as 1) events have been under-reported for decades, especially in developing countries, and 2) losses caused by co-seismic landslide events tend to be classified as secondary losses due to earthquakes.\\
Rapid flow-like landslides, such as rock avalanches and debris flows, show a particularly high hazard potential due to their high mobility, long travel distance and fast propagation speed. In recent years, the geo-hazard community put a lot of effort into developing computational run-out models in order to assess and predict risks associated with rapid landslides and to develop mitigation strategies. Most of the models in practical use are based on a (computationally efficient) 'shallow flow type' process description and depth-averaging techniques, e.g. \citep{Hungr2009a, Pastor2009, Christen2010, Xia2018, Pitman2003}. In these, the flowing material is treated as an 'equivalent fluid' and governed by idealized internal and basal rheologies \cite{Hungr2009a}. Alternative (computationally demanding) models aim at a direct description of fully three-dimensional flow behavior. They hence offer a higher process complexity level, e.g. \citep{Mast2014, Teufelsbauer2011}, yet are typically not feasible for practical hazard mitigation purposes. Detailed reviews of computational run-out models for rapid, flow-like landslide models have been published by \citet{McDougall2017} and \citet{Pastor2018}.\\
An indispensable input to any of these computational landslide run-out models is data that represents the terrain in which the slide is likely to occur. Pioneered by \citet{Miller1958}, digital elevation models (DEMs) have become the most popular form of representing topographies in the scientific community. Methods for generating DEMs have evolved rapidly over decades from conventional approaches like field surveying and topographic map digitizing, to passive and active remote sensing, such as stereoscopic photogrammetry, interferometric synthetic aperture radar (InSAR), and light detection and ranging (LiDAR), see \citet{Wilson2012} for a comprehensive review. Differences between these methods exist in their footprint, cost, resolution and accuracy of the resulting DEM. Whatever method used, however, the resulting DEM will inevitably contain errors that are introduced either during source data acquisition or during data processing. The so-called DEM error hence refers to the difference between the true real world elevations and their DEM representation. Typically, there is a lack of information on the DEM error, which led to notion of 'DEM uncertainty' that refers to what we do not know about the error, see \citet{Wechsler2007}.\\
Nowadays, several global DEM databases, e.g. SRTM \cite{Rodriguez2006}, AW3D30 \cite{Courty2019}, and TanDEM-X \cite{Wessel2018}, as well as some regional DEM databases \cite{Pakoksung2016} are publicly available. Also commercial DEM databases exist that are associated with significant costs, e.g. \citet{Hawker2018}. Online initiatives such as \emph{OpenTopography} facilitate community access and aim at democratizing online availability of high-resolution topography data acquired with LiDAR and other technologies \cite{Krishnan2011}. Despite the broad variety of existing DEM sources, however, we are still facing (and will face in the near future) a very limited availability of high-accuracy DEMs for some regions that are particularly prone to landslide hazards, e.g. in Asia \cite{Froude2018}. Whenever using DEM data for simulation-based landslide hazard analysis, it is hence important to be aware of DEM error and uncertainty, and to consider its potential impact on computational run-out analyses and related computational risk assessments. \\
DEM error has arisen researchers' attention since long. Many efforts have for instance been put into quantifying the error associated with specific DEM sources based on data of higher accuracy, e.g. acquired by satellite measurements \cite{Berry2007, Mouratidis2019}, medium footprint LiDAR \cite{Hofton2006}, or GPS survey \cite{Rodriguez2006, Wessel2018, Bolkas2016, Patel2016, Elkhrachy2018}. Meanwhile, a variety of methods have been devised to classify DEM error into various categories \cite{Oksanen2003, Hengl2004, Fisher2006}. Due to the complexity of potential influencing factors (sensor technology, retrieval algorithms, data processing, land cover and surface morphology, terrain attributes, see \citep{Wilson2012, Fisher2006, Gonga-Saholiariliva2011}), these methods can only constrain the DEM error, and will not deterministically correct for it at all grid points. Hence, DEM uncertainty remains, and has to be accounted for in any subsequent analysis that relies on the DEM data. \\
In this circumstance, a stochastic simulation is an effective computational approach to deal with the situation \cite{Holmes2000}. Instead of considering a single (assumed as accurate) DEM, the fundamental idea of a stochastic simulation in the context of DEM uncertainty propagation is to separate the DEM into a known deterministic contribution and an unknown DEM error. DEM uncertainty is then accounted for by treating the DEM error as a random field consisting of a collection of random variables defined at selected grid points. An ensemble of equiprobable realizations of the random field is then generated based on certain assumptions and available information of DEM error. This could for instance be the so-called root mean square error (RMSE), a minimalistic indicator for the overall error magnitude or a semivariogram that informs about the spatial autocorrelation of the DEM error. Adding the DEM error realizations to the known deterministic DEM contribution results in an ensemble of equiprobable DEM realizations, which can finally be used for a DEM uncertainty propagation analysis. \\
Stochastic simulation methods for DEM uncertainty propagation analyses have been developed since the 1990s and are by now widely applied in many fields, including terrain analysis \cite{Holmes2000, Raaflaub2006, Moawad2018}, flood modeling \cite{Hawker2018, Watson2015, Kiczko2018}, soil erosion modeling \cite{Aziz2012}, landslide susceptibility mapping \cite{Qin2013}, dry block and ash flow modeling \cite{Stefanescu2012}, etc. With respect to rapid, flow-like landslide run-out modeling, very little work has been done to assess the potential impact of DEM uncertainty, most likely due to the complexity, and hence level of sophistication of the associated process models. Meanwhile, however, advances in computing technology led to computationally feasible and well-developed landslide run-out simulation tools. As one of the most important inputs for these tools, a DEM determines the landslide's flow path. A natural next step is hence to consider the impact of DEM uncertainty in these models, as overlooking DEM uncertainty may lead to a bias of risk management decisions in a wrong direction. The major aim of this study is therefore to describe two different approaches in order to incorporate DEM uncertainty into computational landslide run-out analyses, and to investigate and discuss their feasibility, as well as whether DEM uncertainty is critical to landslide run-out and affects its results. \\
This paper is organized as follows: In section~\ref{S:2}, we briefly describe the landslide run-out model used in this study, which is a continuum-mechanical shallow flow model based on the Voellmy-Salm rheology. In section~\ref{S:3}, we recall on various methods to account for DEM uncertainty with a major focus on two approaches, namely an unconditional and a conditional stochastic simulation method. The rest of the paper is devoted to investigating DEM uncertainty propagation for rapid, flow-like landslides based on an integrated workflow that combines the aforementioned computational process model (section~\ref{S:2}) with the stochastic DEM simulations (section~\ref{S:3}). Note, that while in our particular study we chose to use a continuum-mechanical shallow flow process model based on the Voellmy-Salm rheology, the workflow itself is modular and non-intrusive. It would hence also possible to couple the stochastic DEM simulation with any other (DEM based) computational landslide model. Section~\ref{S:4} describes the modular Python-based workflow that we developed in order to set-up and manage the workflow and to interpret its simulation results. We present a series of computational scenarios based upon a historic landslide event in section~\ref{S:5}. All scenarios compare the unconditional and conditional stochastic DEM simulation. Finally, section~\ref{S:6} is devoted to a discussion of our results. Important conclusions are drawn in section~\ref{S:7}.

\section{Landslide process model}
\label{S:2}
Let $\{X, Y, Z\}$ denote a fixed Cartesian coordinate system, in which $X$ and $Y$ are the horizontal axes and $Z$ is the vertical axis. The coordinates of a point in the Cartesian coordinate system are denoted by $(X,Y,Z)$ and a topography can then be expressed as the surface mapping of horizontal $X$ and $Y$ coordinates and represents the elevation at each point, namely $Z(X, Y)$. The mapped topography induces a surface coordinate system $\{x,y,z\}$, in which $x$ and $y$ denote tangential directions and $z$ is perpendicular to the surface.
As mentioned in the introduction, a variety of numerical landslide run-out models have been developed, among these is the family of depth-integrated shallow flow type landslide models. The latter can be further classified based on their applied basal rheology, e.g.  Voellmy, Bingham, Quadratic resistance model, etc., see also \cite{Naef2006, Hungr2009b}. Our study relies on a  Voellmy-Salm (VS) model, hence a depth-averaged continuum mechanical model using the Voellmy basal rheology. Such models are not only applied to rapid landslides, but also to snow avalanches and certain types of rock fall. In our study the process model will be adopted to the historic landslide event in section~\ref{S:5}. The process model along with its underlying assumptions is described in \cite{Bartelt1999,Christen2010}. It assesses the slide's dynamics in terms of flow height and depth-averaged velocity, both of which depend on time $t$ and spatial coordinates $x$ and $y$. They are denoted by $H(x,y,t)$ and $\bm{U}(x,y,t):=(U_x(x,y,t),U_y(x,y,t))^T$ respectively. The governing partial differential equations are derived from first principles of mass and momentum conservation and read

\begin{footnotesize}
\begin{equation}\label{eq:shallowflow}
    \begin{aligned}
\partial_t H + \partial_x(HU_x) + \partial_y(HU_y) &= \dot{Q}(x,y,t)
\\
\partial_t(HU_x) + \partial_x(c_xHU^{2}_{x} + g_zk_{a/p}\frac{H^2}{2}) + \partial_y(HU_xU_y) &= g_xH - \bm{n}_x (\mu g_zH + g{ \left \| \bm{U} \right\|}^2/\xi )
\\
\partial_t(HU_y) + \partial_x(HU_xU_y) + \partial_y(c_yHU^{2}_{y} + g_zk_{a/p}\frac{H^2}{2}) &= g_yH - \bm{n}_y (\mu g_zH + g{ \left \| \bm{U} \right\|}^2/\xi )
 \end{aligned}
\end{equation}
\end{footnotesize}

The first equation of system \eqref{eq:shallowflow} denotes the mass balance, in which $H$ denotes the landslide's height, $U_x$ and $U_y$ its surface tangential velocity components, and $\dot{Q}(x,y,t)$ stands for a mass production source term that accounts for erosion of material along the way. Second and third equations denote the $x$ and $y$ momentum balance, in which $g_x, \; g_y, \; and \; g_z$ are the three components of gravitational acceleration vector $\bm{g}$, $\bm{n}_{x}$ and $\bm{n}_{y}$ are $x$ and $y$ component of the unit vector that opposes the velocity, and $\mu$ and $\xi$ stand for dry Coulomb and 'turbulent' friction coefficients respectively. The two friction parameters are mostly determined by back-analysis based on historic events. Finally, $c_x$ and $c_y$ are velocity shape factors and $k_{a/p}$ denotes the so-called earth-pressure coefficient. Both are difficult to determine for realistic slides, yet have been shown to not critically affect the slide's simulation \cite{Christen2010}. In this study, we therefore set $c_x=c_y=1$ assuming almost plug flow, and $k_{a/p}=1$ since non-hydrostatic internal stresses play a negligible role for flow-like, fully saturated flows \cite{Hungr2005}. 
The topographic surface $Z(X,Y)$ enters the governing equations of the process model implicitly in terms of the spatially varying gravitational acceleration vector $\bm{g}=(g_x,g_y,g_z)^T$. Any error and uncertainty present in the topography representation hence also influences the landslide run-out simulation result.
The VS model had been first proposed to model snow avalanche \cite{Salm1993}. Nowadays, it has been widely applied to other types of gravity-driven rapid mass movements including flow-like landslides \cite{Pastor2018, Frank2015, Hussin2012, Kumar2019}. In this study, the proprietary mass flow simulation platform RAMMS \cite{Christen2010} which provides a GIS integrated implementation of the VS model is used for landslide run-out modeling. It is integrated as a module of our workflow (see section~\ref{S:4}) that is developed for the purpose of DEM uncertainty propagation analysis. For a detailed discussion of the VS model and its implementation, we refer to \cite{Christen2010, Bartelt1999}.

\section{Methods to represent DEM uncertainty}
\label{S:3}
As before in section~\ref{S:2}, we express the topographic surface as a function $Z(X, Y)$ parametrized in horizontal coordinates $X$ and $Y$. In practice, the function $Z(X, Y)$ is often constructed from discrete gridded raster data. We hence assume that a domain of interest $D$ is discretized into the horizontal $X$ and $Y$ direction, which results in a spatial grid defined as 
\begin{equation}
\bm{D}_{mn} = \{D_{ij}=(X_i,Y_j) \ | \ (X_i,Y_j) \in D; \ i=1,2,...,m; \ j=1,2,...,n \}.
\end{equation} 
The elevation data associated with each grid point $D_{ij}$ is defined as
\begin{equation}\bm{Z}_{mn} = \{Z_{ij}=Z(X_{i},Y_{j}) \ | \ \forall \ D_{ij} \in D_{mn} \}.
\end{equation}

The elevation $\bm{Z}_{mn}$ of a common DEM data product might be erroneous with respect to the true values as discussed in the introduction. If we denote the true elevation as \begin{equation}
\bm{Z}^*_{mn} = \{Z^*_{ij}=Z^*(X_{i},Y_{j})\ | \ \forall \ D_{ij} \in D_{mn} \},
\end{equation} the DEM error can be expressed as \begin{equation}
\bm{\epsilon}_{mn} = \{\epsilon_{ij}=Z^*_{ij}-Z_{ij} \ | \ \forall \ D_{ij} \in D_{mn} \}.
\end{equation}

If we knew the error $\bm{\epsilon}_{mn}$, we would be able to recover the real world topographic surface $\bm{Z}^*_{mn}$. The fact, however, that the error is unknown, or we only have limited information about the error implies an uncertainty to the input of our landslide process simulation. Within this study, we will refer to the uncertainty associated to the unknown DEM error as DEM uncertainty. In this circumstance, each $\epsilon_{ij}$ is treated as a random variable and $\bm{\epsilon}_{mn}$ is accordingly treated as a random field, which consists of a collection of random variables $\epsilon_{ij}$. By generating multiple realizations of the random field $\bm{\epsilon}_{mn}$, DEM uncertainty can be represented.  This process is widely known as stochastic simulation. It requires a suitable model to describe the jointed uncertainty of all $\epsilon_{ij}$ based on limited available information of DEM error. The task can be further divided into determining: 
\begin{enumerate}[$\bullet$]
\item the probability distribution function (pdf) of each $\epsilon_{ij}$ which quantifies local uncertainty at each grid point;
\item the correlation between different $\epsilon_{ij}$ which is usually termed as spatial autocorrelation of DEM error. 
\end{enumerate}

According to available information on the DEM error, existing approaches that could be used to solve the two issues can be generally classified into two groups: 
\begin{enumerate}[A)]
\item unconditional stochastic simulation (USS), and 
\item conditional stochastic simulation (CSS).
\end{enumerate}
More specifically, USS is only informed with properties of DEM error, e.g. the RMSE, and thus does not honour any actual DEM error values. CSS is informed with certain number of actual DEM error values at reference locations within the DEM, e.g. obtained from higher accurate reference data, and thus could directly honour the actual DEM error values at reference locations \cite{Fisher2006}.

\subsection{Unconditional stochastic simulation (USS) based on the RMSE}
\label{S:3.1}
Typically available information about the DEM error provided by DEM vendors is the root mean square error (RMSE). For a set of $K$ reference locations, it is defined as

\begin{equation} \label{eq:RMSE}
\text{RMSE} = \sqrt{\frac{1}{K} \sum\limits_{k=1}^{K}(Z^*_{kk}-Z_{kk})^2}.
\end{equation}
Here, $\bm{Z}^*_{KK}=\{Z^*_{kk}=Z^*(X_{k},Y_{k}) \ | \ (X_k,Y_k) \in D; \ k=1,2,...,K \}$ denotes higher accurate elevation values measured at the reference locations and $\bm{Z}_{KK}= \{Z_{kk}=Z(X_{k},Y_{k}) \ | \ (X_k,Y_k) \in D; \ k=1,2,...,K \}$ denotes corresponding elevation values based on the DEM.

It should be noted that while the RMSE is typically available, this is not true for the reference elevation values $\bm{Z}^*_{KK}$ itself. As stated numerous times in the literature, it is critical that the RMSE only provides a global indication of DEM error magnitude without any information about its spatial autocorrelation. Still, it is by far the most widely used DEM error indicator for many DEM databases and mostly the only available information coming along with DEM products. In this circumstance, USS could be used to represent DEM uncertainty and study its propagation into landslide run-out analysis. 

Modeling DEM uncertainty based on USS assumes that all local error values $\epsilon_{ij}$ are independent and fulfill the same univariate Gaussian distribution with a mean ($\mu$) of zero and a standard deviation ($\sigma$) equivalent to the given RMSE. Under this assumption, an ensemble of spatially uncorrelated realizations of the random field $\bm{\epsilon}_{mn}$ can be generated by randomly assigning error values to each $\epsilon_{ij}$ according to its local Gaussian probability distribution. 

In the next step, we have to account for the (unknown) spatial autocorrelation of $\bm{\epsilon}_{mn}$. Potential methods that could be applied are simulated annealing, spatial autoregressive modeling, spatial moving averages, etc., see \cite{Wechsler2007}. Simulated annealing is generally computationally intensive and spatial autoregressive modeling becomes impractical for simulation of large areas \cite{Oksanen2006}. In this study, we use the spatial moving averages method that increases the spatial autocorrelation by filtering spatially uncorrelated realizations with a distance-weighted filter proposed by \citet{Wechsler2006}. For $\epsilon_{ij}$ at one grid point of an uncorrelated realization, its value is replaced by the weighted average of $\epsilon_{ij}$ at all grid points within the filter kernel. The weight decreases with increasing of the distance to the grid point, which is similar to semivariogram trends \cite{Wechsler2006}. The size of the filter denoted as d depends on the maximum autocorrelation length of $\bm{\epsilon}_{mn}$ which again is unknown if the RMSE is the only available information. In practice, d is often determined based on the maximum autocorrelation length of the original DEM \cite{Wechsler2007, Aziz2012}.

Though it relies on some assumptions, such as an appropriate choice of correlation length d, the sketched  approach is generally applicable if RMSE is the only available information. It may become critical if a DEM contains a systematic bias which means that the mean of $\bm{Z}^*_{kk}-\bm{Z}_{kk}$ deviates largely from zero. More specifically, if we follow \cite{Fisher2006, Wessel2018} in defining mean $\mu$ and standard deviation $\sigma$ as

\begin{eqnarray}\label{eq:ME and STD}
\text{$\mu$} = \frac{1}{K} \sum\limits_{k=1}^{K}(Z^*_{kk}-Z_{kk}) 
\quad \text{and} \quad
\text{$\sigma$} = \sqrt{\frac{1}{K-1} \sum\limits_{k=1}^{K}((Z^*_{kk}-Z_{kk})-\text{$\mu$})^2},
\end{eqnarray}
we can express the RMSE in terms of $\mu$ and $\sigma$ as

\begin{equation}\label{eq:ME STD RMSE}
\text{RMSE} = \sqrt{\text{$\mu$}^2+\frac{K-1}{K}\text{$\sigma$}^2}.
\end{equation}

If the number of reference points $K$ is relatively large, $\sqrt{(K-1)/K}$ is close to one. Eq.~(\ref{eq:ME STD RMSE}) then indicates that the RMSE is larger than the standard deviation $\sigma$ if the mean $\mu$ deviates from zero. The difference between the RMSE and $\sigma$ increases with $\mu$ increasing. For example, the $\mu$, $\sigma$, and RMSE of the global TanDEM-X DEM based on about three million reference points are $0.17\,m$, $1.28\,m$, and $1.29\,m$ \cite{Wessel2018}. That of the EU-DEM of Central Macedonia based on 12943 reference points are $1.8\,m$, $3.6\,m$, and $4.0\,m$ while that of the ASTER GDEM of the same area based on the same reference points are $6.8\,m$, $7.6\,m$, and $10.2\,m$ \cite{Mouratidis2019}. This means that assuming the standard deviation of the DEM error to be given as the RMSE consequently overestimates the variability of the DEM error if the mean deviates largely from zero.

The implications of both issues, namely the fact that the filter size d is unknown and has to be subjectively chosen, and that the RMSE provides an insufficient representation of the DEM error, are investigated in the following study. For convenience, the two issues are referred to  as:
\begin{enumerate}
\item[$\bullet$] \emph{unrepresentative RMSE},
\item[$\bullet$] \emph{subjective d}.
\end{enumerate}

\subsection{Conditional stochastic simulation (CSS) based on higher accurate reference data}
\label{S:3.2}
This approach requires the availability of higher accurate reference data at certain reference locations, e.g. from higher accurate DEM products, or GPS surveys. Note, that although these data might be subject to error themselves, it is fair to assume this error to be much smaller. This justifies to use the higher accurate reference data as true elevation values $\bm{Z}^*_{KK}$. Based on $\bm{Z}^*_{KK}$, we could determine the statistics of the DEM error, e.g. the RMSE, the $\mu$ and the $\sigma$ as discussed in section~\ref{S:3.1}. Likewise, we can assess the spatial autocorrelation of the DEM error, e.g. in the form of a semivariogram model (see figure~\ref{fig:semivariogram}). In addition, we know the DEM error at the reference locations exactly, denoted as $\bm{\epsilon}^*_{KK} = \{\ \epsilon^*_{kk} \ |\ k=1,2,...,K \}$. Yet, we still lack knowledge about the DEM error away from the $K$ reference locations, hence the complete random field $\bm{\epsilon}_{mn}$.

In that situation, conditional stochastic simulation (CSS) can be used to simulate, i.e. generate realizations of the random field $\bm{\epsilon}_{mn}$. Many geostatistical methods of conditional simulation could be applied, including sequential simulation algorithms, p-field approach, simulated annealing, etc. \cite{Goovaerts1997}. In this study, we apply a sequential Gaussian simulation. It is the most attractive technique for stochastic spatial simulation  according to \citet{Temme2009} and has been widely utilized in DEM uncertainty propagation analysis \cite{Holmes2000, Aziz2012}. 

The sequential Gaussian simulation sequentially samples each local error $\epsilon_{ij}$ along a random path that consists of all grid points $D_{ij}$ under the multi-Gaussian assumption. This means that assuming the random field $\bm{\epsilon}_{mn}$ to satisfy a multivariate Gaussian distribution, hence each $\epsilon_{ij}$ fulfills a univariate Gaussian distribution denoted as $N(\mu_{ij}, \sigma_{ij})$. The essential idea now is that the mean $\mu_{ij}$ and standard deviation $\sigma_{ij}$ are determined sequentially by means of simple kriging based on: the semivariogram model of DEM error that provides covariances in simple kriging equations, and the conditioning information including $\bm{\epsilon}^*_{KK}$ and previously sampled $\epsilon_{ij}$. By making each univariate Gaussian distribution conditional not only to $\bm{\epsilon}^*_{KK}$ but also to all previously sampled $\epsilon_{ij}$, the semivariogram model of DEM error is reproduced in realizations of $\bm{\epsilon}_{mn}$ \cite{Goovaerts1997}. The process to generate one realization of $\bm{\epsilon}_{mn}$ is as follows:
\begin{enumerate}
    \item[1)] determine a semivariogram model to represent the spatial autocorrelation of DEM error based on normal-score-transformed $\bm{\epsilon}^*_{KK}$;
    \item[2)] define a random path visiting each $D_{ij}$ once;
    \item[3)] at each $D_{ij}$, determine $N(\mu_{ij}, \sigma_{ij})$ using simple kriging based on the semivariogram model and normal-score-transformed $\bm{\epsilon}^*_{KK}$;
    \item[4)] sample a value from $N(\mu_{ij}, \sigma_{ij})$, assign it to $\epsilon_{ij}$, and add $\epsilon_{ij}$ into normal-score-transformed $\bm{\epsilon}^*_{KK}$;
    \item[5)] repeat steps 3) and 4) until all $D_{ij}$ along the path are visited; 
    \item[6)] back-transform all sampled $\epsilon_{ij}$ to the original distribution of $\bm{\epsilon}^*_{KK}$.
\end{enumerate}
Multiple realizations can be generated by defining different random paths.

\section{Implementation}
\label{S:4}
Studying the impact of DEM uncertainty on landslide run-out modeling is computationally intensive and technically demanding. It includes representing DEM uncertainty in terms of a large number of DEM realizations, conducting numerous landslide run-out modeling based on the DEM realizations, and postprocessing extensive output data. In addition, understanding how DEM uncertainty affects terrain attributes may facilitate us to interpret its impact on landslide run-out modeling. This requires the ability to calculate terrain attributes, e.g. slope, ruggedness, etc. of the original DEM as well as the generated DEM realizations.

\begin{figure}[htb]
\centering
\includegraphics[scale=0.8]{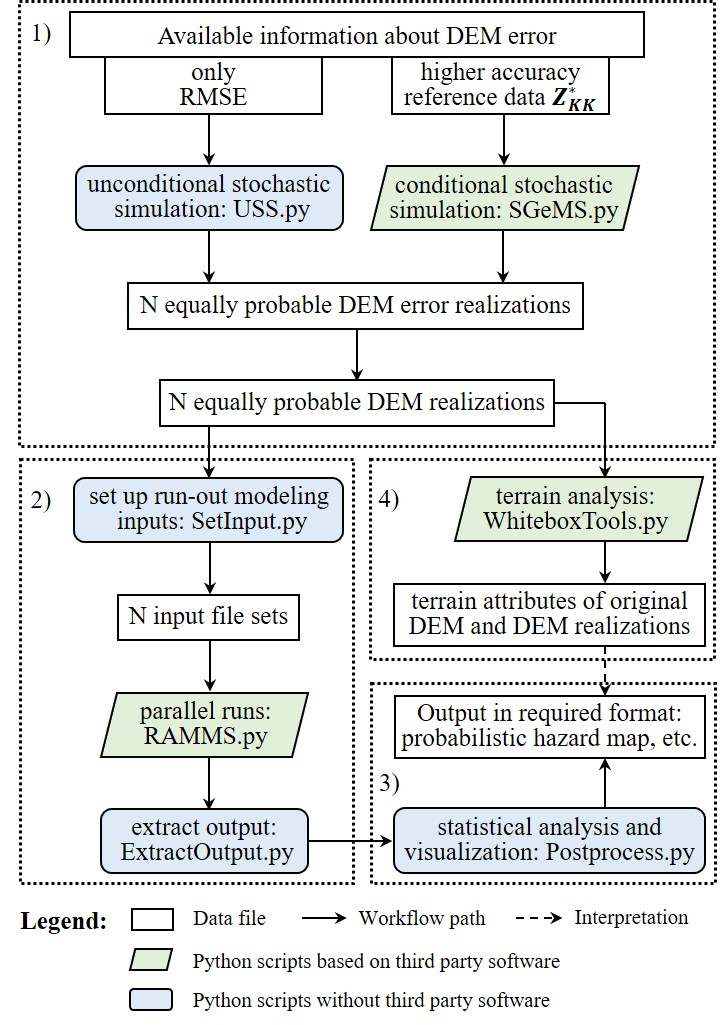}
\caption{Computational workflow of DEM uncertainty propagation in landslide run-out simulation. It is part of our PSI-slide package in development that is designed for the purpose of systematically investigating the impact of various sources of uncertainty on simulating gravity-driven mass movements \cite{Kowalski2018, Zhao2018}. The workflow consists of four modules: 1) DEM uncertainty representation; 2) landslide run-out modeling; 3) statistical analysis and visualization; 4) terrain analysis.}
\label{fig:workflow}
\end{figure}

In this study, we propose a workflow that integrates our own Python implementation of selected aspects of the workflow and existing software as well as toolboxes to solve above mentioned tasks. It is part of our PSI-slide (Predictive Simulation of Slides) package in development that is designed for the purpose of systematically investigating the impact of the various sources of uncertainty on simulating gravity-driven mass movements \cite{Kowalski2018, Zhao2018}. Herein, we focus on DEM uncertainty. Figure~\ref{fig:workflow} illustrates the workflow. It consists of four modules:

\begin{enumerate}
\item[1)] DEM uncertainty representation. In this module, we generate an ensemble of N equally probable DEM realizations to represent DEM uncertainty based on available information about DEM error. USS as introduced in section~\ref{S:3.1} is implemented without third party software (USS.py) for cases in which only the RMSE is available. For cases in which higher accurate reference data is provided, CSS as introduced in section~\ref{S:3.2} is implemented by integrating the sequential Gaussian simulation algorithm of the Stanford Geostatistical Modeling Software (SGeMS) \cite{Remy2009} into our workflow (SGeMS.py). 

\item[2)] Landslide run-out modeling. This module is used to conduct N landslide run-out simulations based on the N DEM realizations generated in module 1). In this study we employ the proprietary mass flow simulation platform RAMMS \cite{Christen2010} which provides a GIS integrated implementation of the VS model. First, a Python script named SetInput.py is called to set up required inputs for each simulation run. Then a Python script named RAMMS.py starts parallel runs of RAMMS using the Python Scoop module. In the end, a Python script named ExtractOutput.py is called to extract user-specified outputs.
\item[3)] Statistical analysis and visualization. This module is used to conduct statistical analysis on the user-specified outputs from module 2) and visualize results. It is mainly based on the Python Numpy and Matplotlib modules. For example, probabilistic hazard map can be produced to indicate potential hazard area.
\item[4)] Terrain analysis. This module is used to analyze terrain characteristics of the original DEM and DEM realizations from module 1), which may help us to interpret outputs from module 3). This is achieved by integrating several terrain analysis tools from WhiteboxTools \cite{Lindsay2018} like calculating slope, aspect, ruggedness index, etc. into our workflow (WhiteboxTools.py).
\end{enumerate}

\section{Case study}
\label{S:5}
This study is based upon a historic landslide and two DEM sources. For the purpose of DEM uncertainty propagation analysis, we assume one DEM source to be more accurate than the other and then obtain higher accurate reference data from the more accurate DEM source to assess elevation error of the less accurate DEM source. We design a series of computational scenarios based on the higher accurate reference data to study the impact of DEM uncertainty on landslide process simulation for both the case when only the RMSE is available and the case when higher accurate reference data is available. Additional computational scenarios are designed to study the \emph{unrepresentative RMSE} and \emph{subjective d} issues as detailed in section~\ref{S:3.1} in the form of a sensitivity analysis. 
\subsection{Scenario background and DEM sources}
\label{S:5.1}
\begin{figure}[htb]
\centering
\includegraphics[width=\linewidth]{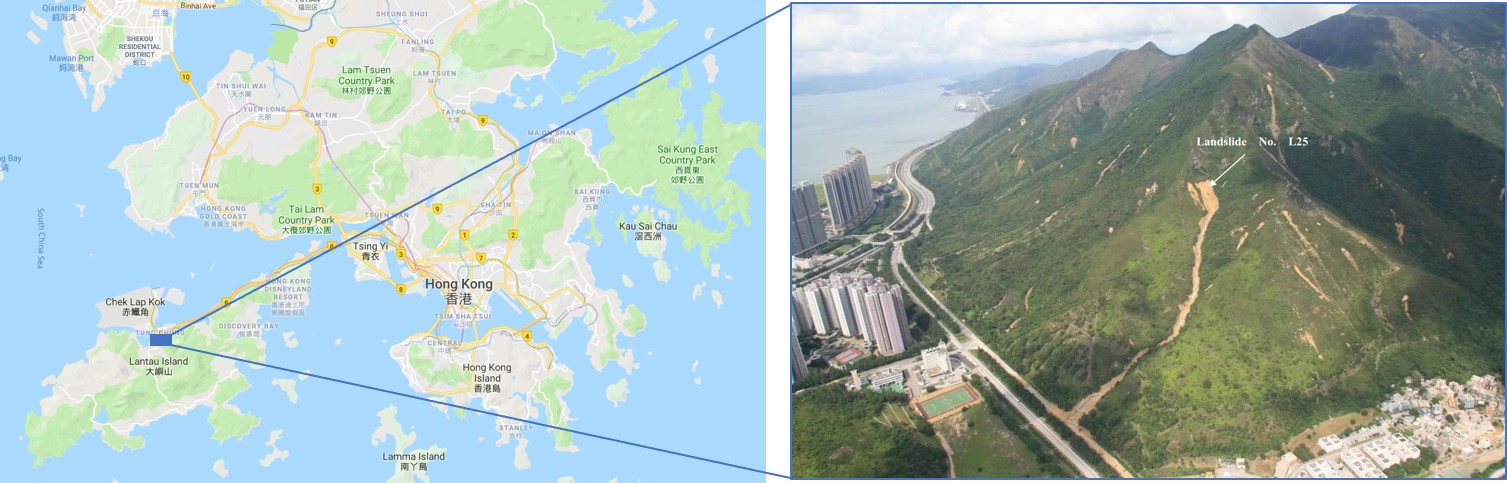}
\caption{The 2008 Yu Tung Road landslide. Left: google map of Hong Kong; right: aerial photograph of Yu Tung Road site after the 2008 landslide. It corresponds to the No. L25 landslide in the GEO report \cite{Report2012}.}
\label{fig:YT landslide}
\end{figure}

The historic landslide happened on June 7 2008 on the hillside above Yu Tung Road in Hong Kong due to an intense rainfall event, see figure~\ref{fig:YT landslide}. It was the largest flow-like landslide out of 19 landslides during that event. Around 3400\,$m^3$ material were mobilized and traveled about 600\,$m$ until deposit. The landslide event had a severe infrastructural impact, as it led to closure of westbound lanes of Yu Tung Road for more than two months \cite{Report2012}. The Yu Tung Road landslide also served as a benchmark case for predictive landslide run-out analysis at the second Joint Technical Committee on Natural Slopes and Landslides (JTC1) Workshop on Triggering and Propagation of Rapid Flow-like Landslides in Hong Kong 2018 \cite{Pastor2018}. 

Two types of DEM data of the Yu Tung Road area had been the basis for this study:
\begin{enumerate}
\item[$\bullet$] A public 5\,$m$ resolution digital terrain model covering the whole area of Hong Kong (HK-DTM). It is downloaded from the website of the survey and mapping office of Hong Kong. The HK-DTM is generated from a series of digital orthophotos, which are derived from aerial photographs taken in 2014 and 2015. The reported accuracy is $\pm5$\,$m$ at $90\%$ confidence level.
\item[$\bullet$] A 2\,$m$ resolution DEM (2m-DEM) covering the main area of the Yu Tung Road landslide event. Its boundary is shown in figure~\ref{fig:reference data}\,(a). It had been provided for the benchmark exercise during the second JTC1 workshop. It was produced based on field mapping after the 2008 Yu Tung Road landslide event.
\end{enumerate}

In this study, we assume the 2m-DEM to be more accurate than the 5\,$m$ resolution HK-DTM. Similar to our consideration at the beginning of section~\ref{S:3.2}, the 2m-DEM and 5\,$m$ resolution HK-DTM correspond to $\bm{Z}^*$ and $\bm{Z}$ as defined in section~\ref{S:3}. A set of higher accurate reference data $\bm{Z}^*_{KK}$ can hence be determined to provide information to represent uncertainty of the 5\,$m$ resolution HK-DTM.

It should be noted that the 2m-DEM and 5\,$m$ resolution HK-DTM were produced in different time periods. After the 2008 Yu Tung Road landslide, debris-resisting barriers and a road had been built in the area within the red circle and blue rectangle in figure~\ref{fig:reference data}\,(a) respectively. They are reflected in the HK-DTM but not in the 2m-DEM, which leads to large inconsistency between the two DEMs in that area. Therefore, to avoid unrealistically large error of the HK-DTM, data from the 2m-DEM in that area is excluded from higher accurate reference data $\bm{Z}^*_{KK}$.

\subsection{DEM realizations}
\label{S:5.2}
\subsubsection{Information of DEM error}
\begin{figure}[htb]
\centering
\includegraphics[width=\linewidth]{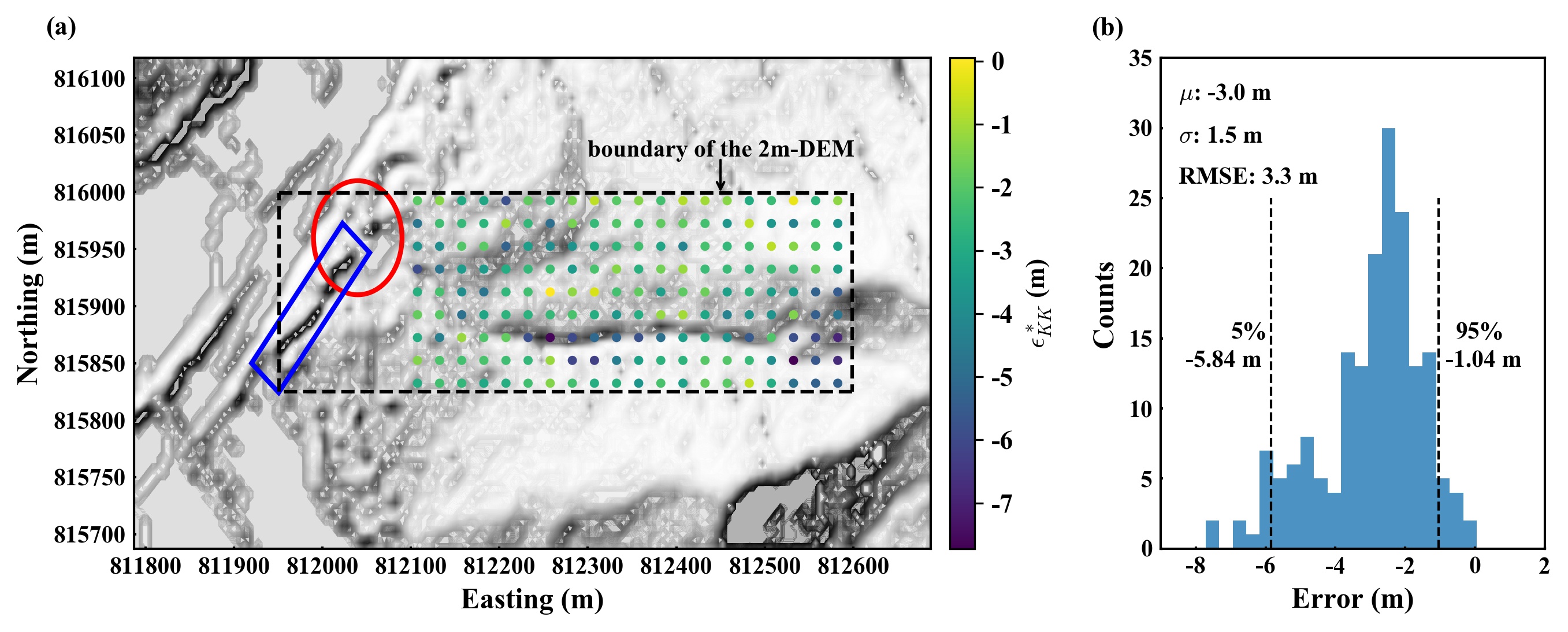}
\caption{(a) Elevation error $\bm{\epsilon}^*_{KK}\{K=180\}$ of the HK-DTM at 180 reference locations. The background is the hillshade plot of the HK-DTM. Debris-resisting barriers and a road in the circle and rectangle area constructed after the 2008 landslide event are represented in the HK-DTM but not in the 2m-DEM. It causes inconsistency between the two DEMs in that area. To avoid unrealistically large error of the HK-DTM, data from the 2m-DEM in that area is excluded from higher accurate reference data. (b) Histogram of $\bm{\epsilon}^*_{KK}\{K=180\}$.  The RMSE is larger than the standard deviation ($\sigma$) since the mean ($\mu$) is not zero. As discussed in section~\ref{S:3.1}, this indicates that assuming the standard deviation of the HK-DTM error being equivalent to the RMSE in USS would overestimate the variability of the HK-DTM error.}
\label{fig:reference data}
\end{figure}

As shown in figure~\ref{fig:reference data}\,(a), we evenly pick 180 reference locations from the HK-DTM grid points within the boundary of the 2m-DEM. Higher accurate reference data at these locations is obtained from the 2-m DEM using bilinear interpolation, denoted as $\bm{Z}^*_{KK}\{K=180\}$. Subtracting the corresponding elevation values of the HK-DTM $\bm{Z}_{KK}\{K=180\}$ from $\bm{Z}^*_{KK}\{K=180\}$, we obtain elevation error values of the HK-DTM at the 180 reference locations, denoted as $\bm{\epsilon}^*_{KK}\{K=180\}$.

\begin{figure}[htb]
\centering
\includegraphics[scale=0.8]{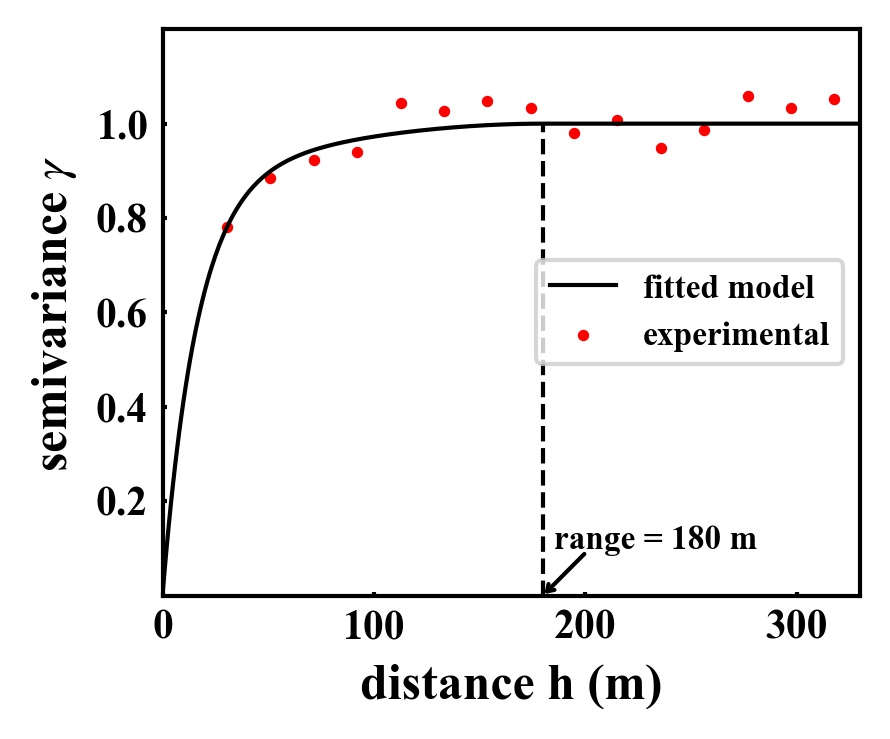}
\caption{Experimental semivariances based on $\bm{\epsilon}^*_{KK}\{K=180\}$ and fitted parametrized semivariogram model given by eq.~(\ref{eq:semivariogram}). The range of the semivariogram model is 180\,$m$. It indicates the maximum autocorrelation length of DEM error, on which the size of the spaial moving filter d depends (see section~\ref{S:3.1}).}
\label{fig:semivariogram}
\end{figure}

Figure~\ref{fig:reference data}\,(b) shows the histogram of $\bm{\epsilon}^*_{KK}\{K=180\}$. $90\%$ of the elevation error values are within -5.84\,$m$ and $-$1.04\,$m$, which is close to the reported accuracy (see section~\ref{S:5.1}). The $\mu$, $\sigma$, and RMSE according to eq.~(\ref{eq:ME and STD}) and eq.~(\ref{eq:RMSE}) are -3.0\,$m$, 1.5\,$m$, and 3.3\,$m$ respectively. Here, it should be noted that the RMSE is larger than the $\sigma$ since the $\mu$ is not zero which indicates a systematic bias. As discussed in section~\ref{S:3.1}, this also indicates that assuming the standard deviation of the HK-DTM error being equivalent to the RMSE in USS would overestimate the variability of the HK-DTM error.

Based on $\bm{\epsilon}^*_{KK}\{K=180\}$, we can determine an isotropic semivariogram model which describes the spatial autocorrelation of the HK-DTM error. It results in
\begin{equation} \label{eq:semivariogram}
\gamma(h) = 0.1 \times Sph(\frac{h}{180}) + 0.9 \times Exp(\frac{h}{50}).
\end{equation}

Here, $Sph(\cdot)$ and $Exp(\cdot)$ denote the basic spherical and exponential semivariogram models \cite{Goovaerts1997} and $h$ denotes the horizontal distance between any two locations. A comparison between the experimental semivariance values based on $\bm{\epsilon}^*_{KK}\{K=180\}$ and the parametrized semivariogram model given by eq.~(\ref{eq:semivariogram}) can be seen in figure~\ref{fig:semivariogram}. Semivariance is a measure of spatial dependence between DEM error values at two different locations. The continuous semivariogram model is fitted to the experimental semivariance values so as to deduce semivariance values for any possible distance $h$ required by simple kriging \cite{Goovaerts1997}. The range of the semivariogram model is 180\,$m$. It indicates the maximum autocorrelation length of the HK-DTM error, on which the size of the spatial moving filter d depends (see section~\ref{S:3.1}).

\subsubsection{DEM uncertainty scenarios}
\label{S:5.2.2}
As mentioned in section~\ref{S:3}, DEM users are often restricted to DEM error information in the form of a single RMSE value per data product. Rarely, they have higher accurate reference data. In order to account for both situations, two corresponding 'information levels' are considered in the following study.
\begin{enumerate}
\item[A)] Rudimentary error information: the RMSE only. In this situation, the RMSE is assumed to be the only available error information of the 5\,$m$ resolution HK-DTM. In order to compare results to B), we employ the RMSE 3.3\,$m$ as generated based on $\bm{Z}^*_{KK}\{K=180\}$, as well as the size of the spatial moving filter d 180\,$m$ to match the range of the fitted semivariogram model in figure~\ref{fig:semivariogram}. USS introduced in section~\ref{S:3.1} is used to generate N~realizations of the HK-DTM, denoted as $\text{USS}_\text{N}$\{RMSE=3.3, d=180\}.
\item[B)] Highly informed: higher accurate reference data. In this situation, $\bm{Z}^*_{KK}\{K=180\}$ is assumed to be available. That means we know the error $\bm{\epsilon}^*_{KK}\{K=180\}$ at the reference locations exactly and the fitted semivariogram model based on $\bm{\epsilon}^*_{KK}\{K=180\}$. CSS introduced in section~\ref{S:3.2} is used to generate N~realizations of the HK-DTM, denoted as $\text{CSS}_\text{N}$.
\end{enumerate}

Following the two nominal scenarios A) and B) that are based on specific error $\bm{\epsilon}^*_{KK}\{K=180\}$ at  reference locations determined from the available data sources, we also want to analyze the impact of \emph{unrepresentative RMSE} and \emph{subjective d} issues of USS as introduced in section~\ref{S:3.1} in the form of a sensitivity analysis. Hence, to what extent can we trust the results of USS if only a single RMSE value per data product is available. Additional three values of the RMSE that are 0.5\,$m$, 1.5\,$m$, and 2.5\,$m$ with a fixed d 180\,$m$ are used as inputs for USS to study the \emph{unrepresentative RMSE} issue. It should be noted that the RMSE 1.5\,$m$ corresponds to the 'true' standard deviation $\sigma$ based on $\bm{\epsilon}^*_{KK}\{K=180\}$, see figure~\ref{fig:reference data}\,(b). Another additional three values of d that are 0\,$m$, 90\,$m$, and 270\,$m$ with a fixed RMSE 3.3\,$m$ are used to consider the \emph{subjective d} issue. The corresponding realizations of the HK-DTM are denoted as $\text{USS}_\text{N}$\{RMSE=0.5, 1.5, 2.5, d=180\} and $\text{USS}_\text{N}$\{RMSE=3.3, d=0, 90, 270\}.

To sum up, all the scenarios for stochastic simulation are listed in table~\ref{tab:scenarios}.

\begin{table}[htbp]
\centering 
\caption{Scenarios for stochastic simulation}\label{tab:scenarios}
\begin{tabular}{@{} lcc @{}}
\Xhline{2\arrayrulewidth}
Method & Input ($m$) & Source/Purpose \\
\Xhline{2\arrayrulewidth}
\multirow{3}{4em}{USS} & RMSE=3.3, d=180 & $\bm{Z}^*_{KK}\{K=180\}$ \\ 
& RMSE=0.5, 1.5, 2.5, d=180 & \emph{unrepresentative RMSE} \\ 
& RMSE=3.3, d=0, 90, 270 & \emph{subjective d} \\ 
{CSS} & semivariogram, $\bm{\epsilon}^*_{KK}\{K=180\}$ & $\bm{Z}^*_{KK}\{K=180\}$ \\
\Xhline{2\arrayrulewidth}
\end{tabular}
\end{table}

\subsubsection{Number of DEM realizations}

\begin{figure}[htb]
\centering
\includegraphics[width=\linewidth]{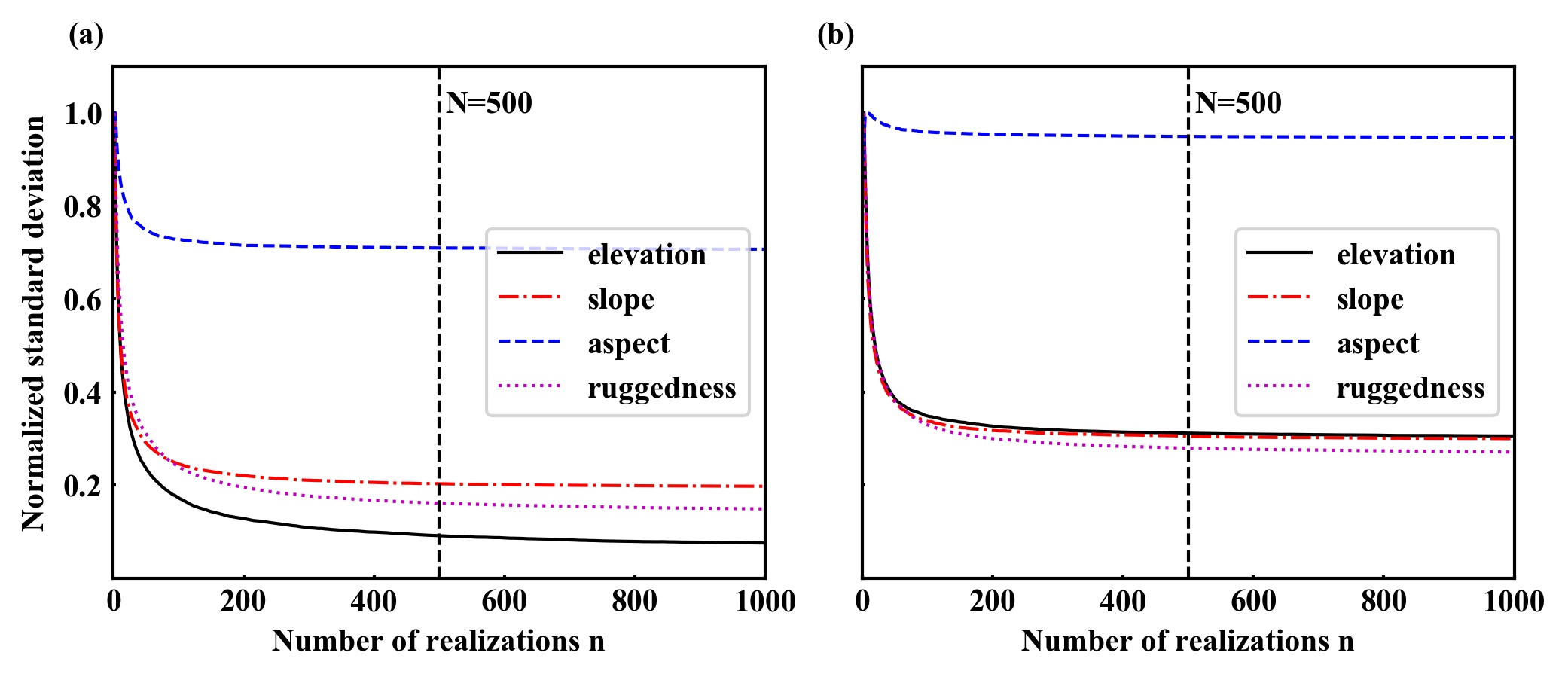}
\caption{The relative change of topographic attributes with respect to the number of HK-DTM realizations. The realizations are generated with (a) $\text{USS}_\text{N}$\{RMSE=3.3, d=180\}; (b) $\text{CSS}_\text{N}$. Beyond N=500, adding more realizations has little impact on topographic attributes. Therefore, we set N=500 for all computational scenarios in table~\ref{tab:scenarios}.}
\label{fig:std of std}
\end{figure}

The integrity of a stochastic simulation requires a large number of DEM realizations, while more realizations naturally take many computational resources. Thus one has to find a reasonable compromise. Typically, this can be found through a representative convergence study. Since in our study we address the impact of topographic uncertainty on landslide run-out simulation, we analyze the relative change of topographic attributes with an increasing number of HK-DTM realizations in a preliminary study. Herein, 1000 HK-DTM realizations are generated for the two 'information levels' A) and B) as introduced in section~\ref{S:5.2.2} respectively, namely $\text{USS}_\text{N=1000}$\{RMSE=3.3, d=180\} and $\text{CSS}_\text{N=1000}$. Topographic attributes including slope, aspect, and ruggedness at all HK-DTM grid points are calculated for each realization.

We define an indicator of the relative change similarly as in \cite{Raaflaub2006} to investigate the converging behaviour. Taking slope as an example, for a given number $n$ of HK-DTM realizations, we first calculate the standard deviation of slope at each HK-DTM gird point over the $n$ realizations. The calculated standard deviation values at all grid points constitute a grid of standard deviation values. Then we calculate the standard deviation of the grid of standard deviation values, which leads to a single standard deviation value for the given number $n$. For each $n$ from 1 to 1000, we can correspondingly calculate a standard deviation value. The same procedure is applied to aspect, ruggedness, and elevation.

Figure~\ref{fig:std of std} shows plots of normalized standard deviation of the grid of standard deviation values with respect to the number of HK-DTM realizations for the two situations A) and B). It can be seen that for situation A), aspect levels out first, followed by slope, ruggedness, and elevation. Beyond 500 realizations, there is little change of normalized standard deviations. This indicates adding more realizations has little impact on topographic attributes. For situation B), aspect also levels out first while the rest three show less difference. Compared to A), B) converges faster which indicates CSS requires less DEM realizations than USS. Nevertheless, we set N=500 for the remainder of this study both for USS and CSS. We generate 500 HK-DTM realizations for each scenario input set as listed in table~\ref{tab:scenarios}.

\begin{figure}[htb]
\centering
\includegraphics[width=\linewidth]{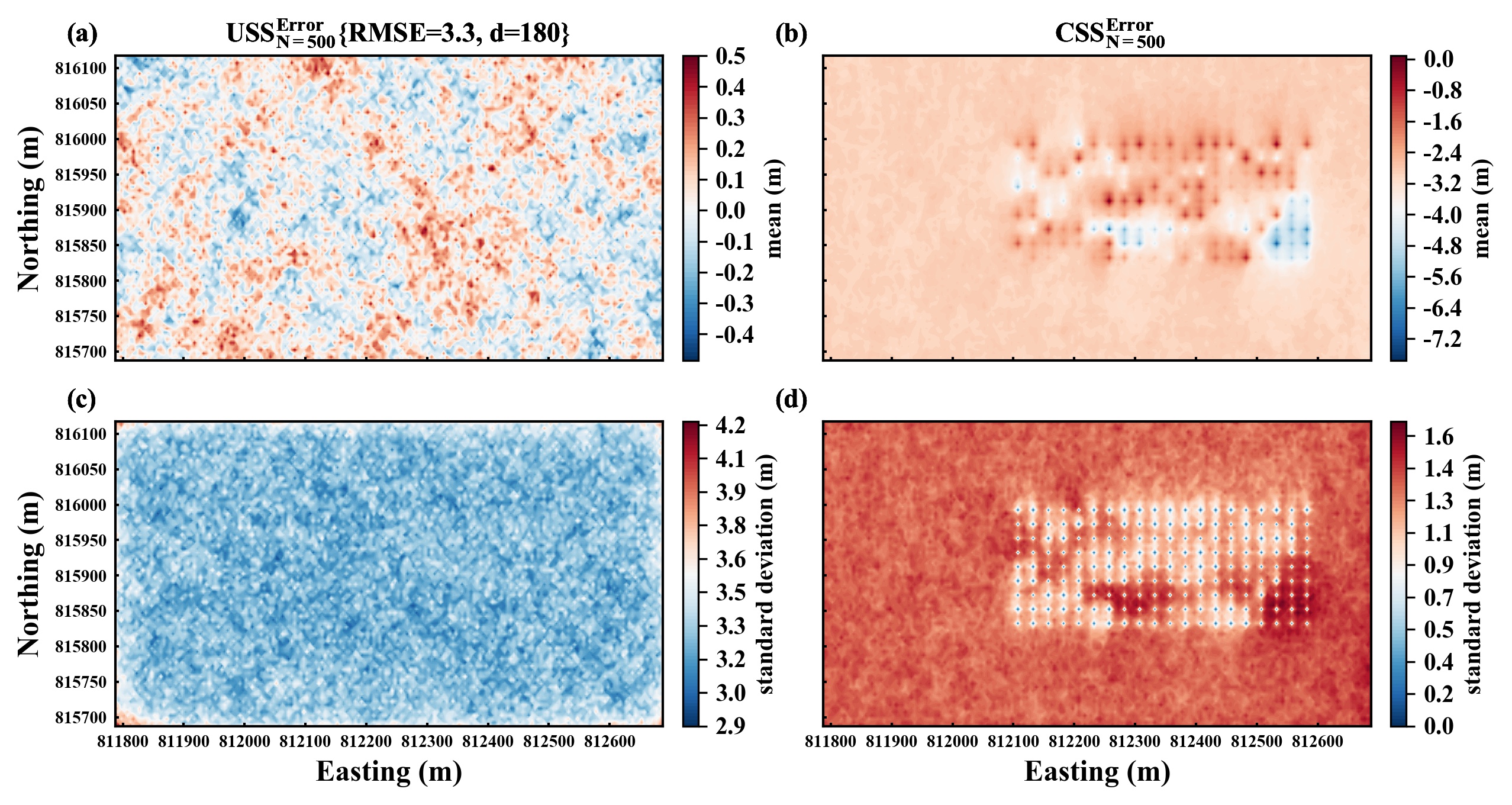}
\caption{Statistics of HK-DTM error realizations. (a) mean and (c) standard deviation grid of $\text{USS}^\text{Error}_\text{N=500}$\{RMSE=3.3, d=180\}. The mean and standard deviation values are centered around 0\,$m$ and 3.3\,$m$; (b) mean and (d) standard deviation grid of $\text{CSS}^\text{Error}_\text{N=500}$. The mean values at grid points away from the reference locations are centered around the mean ($\mu$) $-3.0\,m$ of $\bm{\epsilon}^*_{KK}\{K=180\}$ and are equal to $\bm{\epsilon}^*_{KK}\{K=180\}$ at the reference locations. The standard deviation values at grid points away from the reference locations are centered around the standard deviation ($\sigma$) $1.5\,m$ of $\bm{\epsilon}^*_{KK}\{K=180\}$ and vanish at the reference locations. This matches the assumptions underlying USS and CSS as introduced in section~\ref{S:3}.}
\label{fig:mean and std grid}
\end{figure}
\subsubsection{Statistics of DEM error realizations}
In order to conduct a further quality check of our implementation of both USS and CSS, we investigate the corresponding DEM error realizations of the $\text{USS}_\text{N=500}$\{RMSE=3.3, d=180\} and $\text{CSS}_\text{N=500}$ scenarios, denoted as $\text{USS}^\text{Error}_\text{N=500}$ \{RMSE=3.3, d=180\} and $\text{CSS}^\text{Error}_\text{N=500}$ respectively. Ideally, the local mean $\mu_{ij}$ and standard deviation $\sigma_{ij}$ of DEM error realizations at each grid point $D_{ij}$ should match the underlying assumptions as introduced in section~\ref{S:3} if the number of DEM error realizations is sufficiently large.

Figure~\ref{fig:mean and std grid}\,(a) and (c) show the mean and standard deviation grid of the $\text{USS}^\text{Error}_\text{N=500}$\{RMSE=3.3, d=180\}. It can be seen that the mean values at all grid points are centered around 0\,$m$. The standard deviation values are centered around 3.3\,$m$. This corresponds to the assumption underlying USS that all $\epsilon_{ij}$ fulfill a same univariate Gaussian distribution with a mean ($\mu$) of zero and a standard deviation ($\sigma$) given by the RMSE (see section~\ref{S:3.1}).

Figure~\ref{fig:mean and std grid} \,(b) and (d) show the mean and standard deviation grid of the $\text{CSS}^\text{Error}_\text{N=500}$. The mean values at grid points away from the reference locations are centered around the mean ($\mu$) $-3.0\,m$ based on $\bm{\epsilon}^*_{KK}\{K=180\}$. They become close to $\bm{\epsilon}^*_{KK}\{K=180\}$ with the decrease of distance between grid points and the reference locations, and are equal to $\bm{\epsilon}^*_{KK}\{K=180\}$ at the reference locations. Similarly, the standard deviation values at grid points away from the reference locations are centered around the standard deviation ($\sigma$) $1.5\,m$ based on $\bm{\epsilon}^*_{KK}\{K=180\}$. They vanish at the reference locations. This also corresponds to the assumption underlying CSS that each $\epsilon_{ij}$ fulfill a univariate Gaussian distribution with a mean $\mu_{ij}$ and standard deviation $\sigma_{ij}$ given by the simple kriging estimate and simple kriging standard deviation at $D_{ij}$ (see section~\ref{S:3.2}).

\subsection{Landslide process simulation setup}
With the DEM realizations generated in section~\ref{S:5.2}, we can study the impact of DEM uncertainty on landslide process simulation. Here, we introduce the key inputs and our setup for the process simulation. 
\subsubsection{Model input}
Release zone area and fracture height, friction parameters, and a DEM are three key inputs for performing a deterministic landslide process simulation based on the VS model and utilizing the mass movement simulation platform RAMMS \cite{Christen2010}. For all scenarios, we consistently use the release zone area as provided for the benchmark exercise during the second JTC1 workshop, which match that of the 2008 Yu Tung Road landslide \cite{Pastor2018} as shown in figure~\ref{fig:deterministic}\,(b). The fracture height is assumed to be 1.2\,$m$ leading to a release volume of around 2900\,$m^3$ based on the 5\,$m$ resolution HK-DTM. The friction parameters $\mu$ and $\xi$ used in this study are 0.105 and 300\,$m/s^2$ respectively. They are suggested in the GEO report issued by Geotechnical Engineering Office of Hong Kong, which are obtained using back-analysis with information from a video capturing the lower portion of the landslide and detailed field mapping after the landslide \cite{Report2012}. The HK-DTM and all HK-DTM realizations generated in section~\ref{S:5.2} are used as DEM inputs. Entrainment is not considered in this study.

\subsubsection{Simulation ensembles}
\label{S:5.3.2}
We denote a deterministic landslide process simulation based on a DEM as a simulation run and N deterministic landslide process simulations based on N DEM realizations as a simulation ensemble. The following deterministic simulation and simulation ensembles are conducted based on the original HK-DTM and the aforementioned computational scenarios, see table~\ref{tab:scenarios}. They are named after the corresponding DEM and DEM realizations.
\begin{enumerate}
\item[1)] Deterministic simulation HK-DTM: one landslide process simulation run is conducted based on the original HK-DTM. This one time simulation corresponds to, what is traditionally done in a simulation based hazard assessment study. The results serve as the basis to assess the impact of DEM uncertainty.
\item[2)] $\text{USS}_\text{N=500}$\{RMSE=3.3, d=180\} ensemble: 500 landslide process simulations are conducted based on the $\text{USS}_\text{N=500}$\{RMSE=3.3, d=180\} DEM realizations as introduced in section~\ref{S:5.2}. Each of them is referred to as  $\text{USS}^n_\text{N=500}$\{RMSE=3.3, d=180\}, with $n=1, 2, ..., 500$. This ensemble allows us to access the impact of DEM uncertainty if only the RMSE is available.
\item[3)] $\text{CSS}_\text{N=500}$ ensemble: 500 landslide process simulations are conducted based on the $\text{CSS}_\text{N=500}$ DEM realizations. Similar to 2), each of them is referred to as $\text{CSS}^n_\text{N=500}$ with $n=1, 2, ..., 500$. This ensemble allows us to assess the impact of DEM uncertainty if higher accurate reference data is available.
\item[4)] $\text{USS}_\text{N=500}$\{RMSE=0.5, 1.5, 2.5, d=180\} ensembles: 500 landslide process simulations are conducted for three different RMSE values respectively while keeping the maximum autocorrelation length d constant. They lead to 1500 process simulations. The results allow us to discuss the \emph{unrepresentative RMSE} issue as detailed in section~\ref{S:3.1}. They can be also used to discuss the relationship between the degree of DEM uncertainty and its impact.
\item[5)] $\text{USS}_\text{N=500}$\{RMSE=3.3, d=0, 90, 270\} ensembles: 500 landslide process simulations are conducted for three different maximum autocorrelation length values respectively while keeping the RMSE constant. They lead to 1500 process simulations. The results allow us to discuss the \emph{subject d} issue as detailed in section~\ref{S:3.1}. 
\end{enumerate}
All in all this adds up to one deterministic simulation run HK-DTM, as well as to simulation ensembles 500 process simulations each, the $\text{USS}_\text{N=500}$ \{RMSE=3.3, d=180\} ensemble and $\text{CSS}_\text{N=500}$ ensemble, that are constructed from higher accurate reference data based on the 2m-DEM, as well as 3000 additional process simulations to result in six ensembles $\text{USS}_\text{N=500}$\{RMSE=0.5, 1.5, 2.5, d=180\} and $\text{USS}_\text{N=500}$\{RMSE=3.3, d=0, 90, 270\} to test sensitivities. Each process simulation takes around one minute on a laptop with Intel Core i7-9750H CPU, adding up to around 67 hours simulation time.

\section{Results and discussions}
\label{S:6}
This section is organized according to the simulation ensembles introduced in section~\ref{S:5.3.2}. Section~\ref{S:6.1} presents the results of the deterministic simulation HK-DTM which serves as the basis for all following discussions. Section~\ref{S:6.2} is devoted to analyze the impact of DEM uncertainty on landslide process simulation in the case of RMSE only ($\text{USS}_\text{N=500}$\{RMSE=3.3, d=180\} ensemble) and available higher accurate reference data ($\text{CSS}_\text{N=500}$ ensemble). In section~\ref{S:6.3}, the \emph{unrepresentative RMSE} and \emph{subjective d} issues are discussed based on the ensembles described in section~\ref{S:5.3.2} 4) and 5).

\subsection{Deterministic simulation HK-DTM}
\label{S:6.1}
In a continuum mechanical landslide process model such as used for this study and introduced in section~\ref{S:2}, the landslide flow behaviour is characterised by its spatially varying height and velocity distribution over time, denoted as $H(x,y,t)$ and $\bm{U}(x,y,t)$. For the purpose of landslide hazard assessment and mitigation measure development, hence maximum height and velocity data through the duration of the landslide are most informative. Thus, we focus on the maximum values of $H(x,y,t)$ and $\bm{U}(x,y,t)$ over all time, denoted as $H_\text{max}(x,y):=\{\text{max}\,H(x,y,t) \ | \ t \in [t_{mobilized},t_{deposit}]\}$ and $\left \| \bm{U}_\text{max}(x,y) \right\|:=\{\text{max}\left \| \bm{U}(x,y,t) \right\| \ | \ t \in [t_{mobilized},t_{deposit}]\}$.

\begin{figure}[htb]
\centering
\includegraphics[scale=0.7]{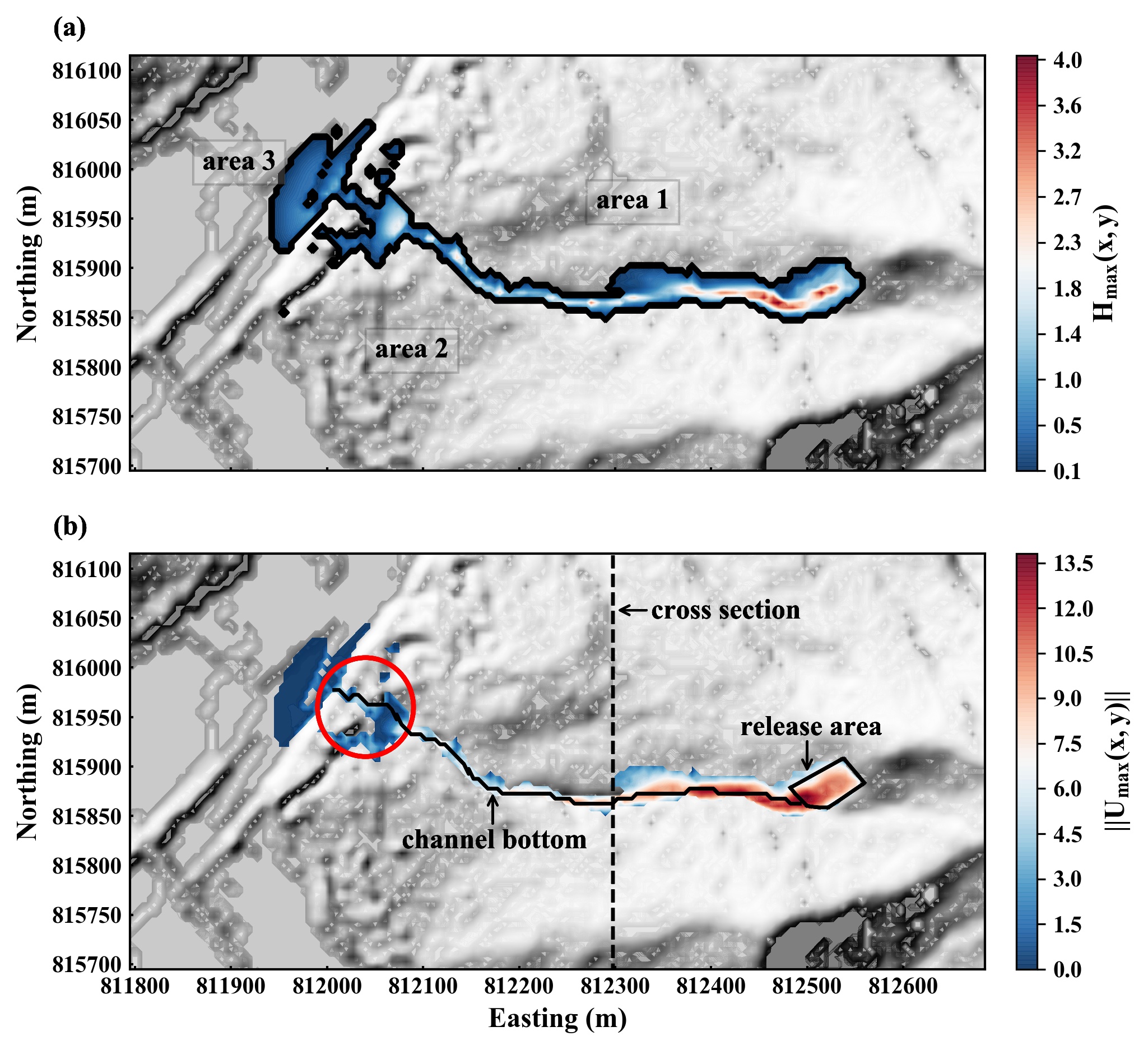}
\caption{Results of the deterministic simulation HK-DTM. (a) $H_\text{max}(x,y)$ above a cut-off threshold of 0.1\,$m$. The black outline is the 0.1\,$m$ contour of $H_\text{max}(x,y)$. The area within this outline is regarded as hazard area. Area~1-3 are denoted for latter discussions. (b) $\left \| \bm{U}_\text{max}(x,y) \right\|$ above a cut-off threshold of 0.01\,$m/s$. The relatively high elevation area within the red circle decelerates and holds back the flow material. The channel bottom and cross section are denoted for latter discussions.}
\label{fig:deterministic}
\end{figure}

Figure~\ref{fig:deterministic}\,(a) and (b) show $H_\text{max}(x,y)$ and $\left \| \bm{U}_\text{max}(x,y) \right\|$ as given by the deterministic simulation HK-DTM. It should be noted that there is a relatively high elevation area at the end part of the channel in the HK-DTM as denoted within the red circle in figure~\ref{fig:deterministic}\,(b). It corresponds to the construction of debris-resisting barriers after the 2008 Yu Tung Road landslide as introduced in section~\ref{S:5.1}. The flow material is decelerated and held back here. We will come back to this point latter in section~\ref{S:6.2.1}.

Landslide run-out distance is often characterised in terms of its apparent friction angle. The tangent of the apparent friction angle is equal to the ratio of the landslide fall height and the run-out distance \cite{DeBlasio2008}. The apparent friction angle evaluated from the deterministic simulation is $16.80^{\circ}$. These results are used as reference to assess the impact of DEM uncertainty in the following discussions.

\clearpage
\subsection{$\text{USS}_\text{N=500}$\{RMSE=3.3, d=180\} ensemble and $\text{CSS}_\text{N=500}$ ensemble}
\label{S:6.2}
While it is straightforward to present results of a deterministic simulation run as shown in section~\ref{S:6.1}, a stochastic simulation based ensembles of N simulation run call for tailored statistic to manage and interpret the extensive output data. First, we define the hazard probability $P_{(x_l,y_l)}$ at one location  $(x_l,y_l)$ as the frequency of $H_\text{max}(x_l,y_l)$ exceeding a certain pre-defined height threshold value, hence
\begin{equation} \label{eq:hazardprob}
P_{(x_l,y_l)}=\frac{\sum_{n=1}^{N}P^n_{(x_l,y_l)}}{N}, \: P^n_{(x_l,y_l)}=\begin{cases}
    1, & \text{if $H^n_\text{max}(x_l,y_l) \geq$ \,threshold}\\
    0, & \text{otherwise}
  \end{cases}
\end{equation}
where $H^n_\text{max}(x_l,y_l)$ denotes the maximum flow height at location $(x_l,y_l)$ for the n-th simulation run of the corresponding ensemble. $P^n_{(x_l,y_l)}$ hence informs whether location $(x_l,y_l)$ is within the hazard area of the n-th simulation run for a given threshold, and $P_{(x_l,y_l)}$ about the resulting hazard probability at location $(x_l,y_l)$ considering the complete ensemble. Here, the threshold is set as 0.1\,$m$ which matches the cut-off threshold of the deterministic simulation HK-DTM in figure~\ref{fig:deterministic}\,(a). Evaluation of hazard probabilities at all locations then gives rise to a probabilistic hazard map \cite{Stefanescu2012}, which provides an overall view of the DEM uncertainty impact.

Besides assessing the overall impact of DEM uncertainty in terms of the probabilistic hazard map, we will also discuss the local impact of DEM uncertainty on dynamic flow properties, focusing on $H_\text{max}(x,y)$ and $\left \| \bm{U}_\text{max}(x,y) \right\|$ at locations along the channel bottom and the channel cross section denoted in figure~\ref{fig:deterministic}\,(b).

\subsubsection{Probabilistic hazard maps}
\label{S:6.2.1}

\begin{figure}[htbp]
\centering
\includegraphics[width=\linewidth]{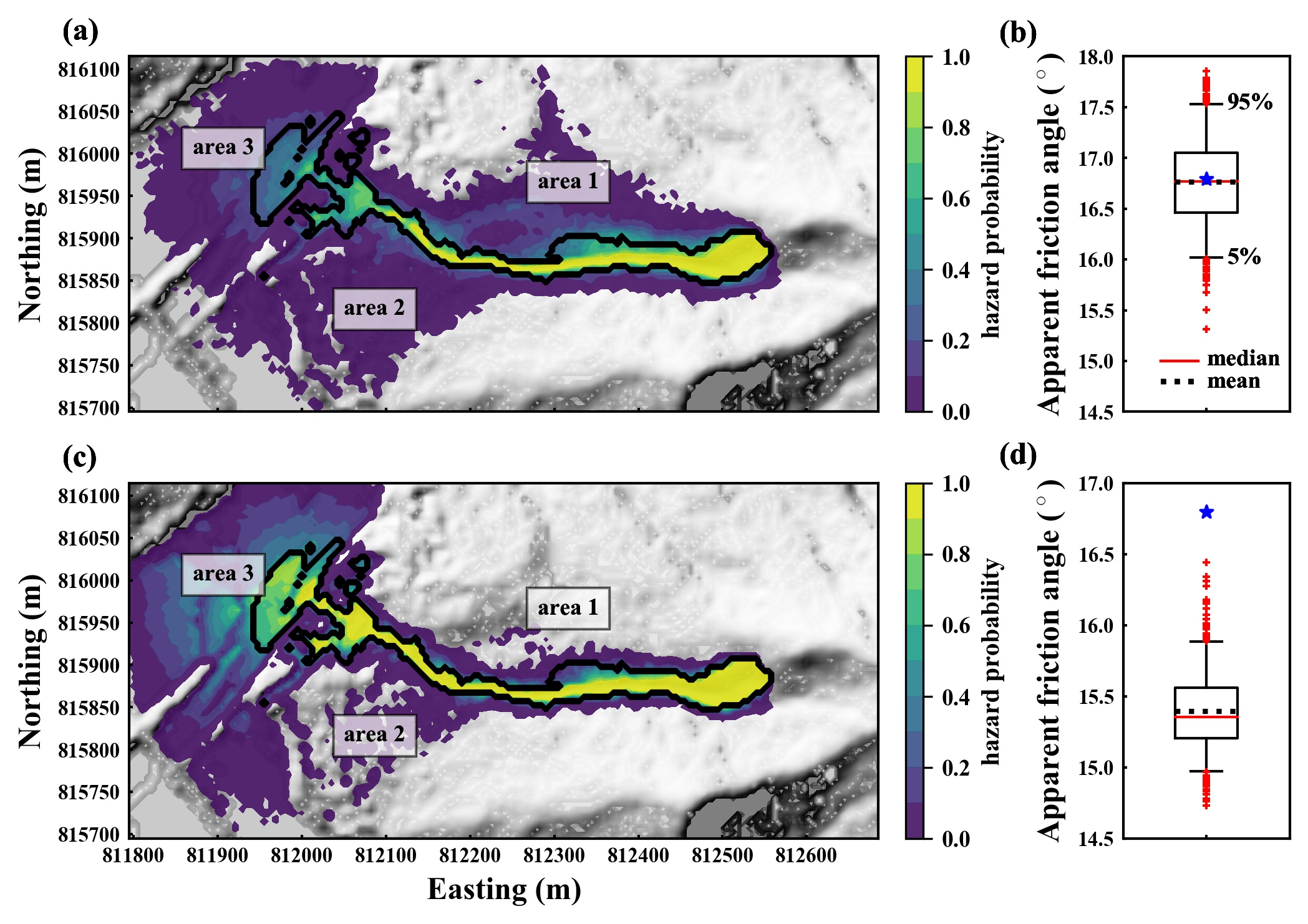}
\caption{(a) Probabilistic hazard map and (b) corresponding apparent friction angle distribution of $\text{USS}_\text{N=500}$ \{RMSE=3.3, d=180\} ensemble; (c) probabilistic hazard map and (d) corresponding apparent friction angle distribution of $\text{CSS}_\text{N=500}$ ensemble. The black outline plotted on the hazard maps represents the deterministic hazard area (see figure~\ref{fig:deterministic}\,(a)). In the boxplots, the blue star denotes the apparent friction angle of the deterministic simulation HK-DTM (see section~\ref{S:6.1}). The difference between the deterministic and the ensemble based hazard area is most pronounced in area~1-3 for $\text{USS}_\text{N=500}$\,\{RMSE=3.3, d=180\} ensemble and in area~3 for $\text{CSS}_\text{N=500}$ ensemble. Our main findings are: 1) accounting for DEM uncertainty may significantly increase the potential hazard area; 2) the potential hazard area is highly related to the variability of DEM error and topographic details of the original DEM; 3) USS based on the RMSE only may overestimate the spread of potential hazard area and underestimate the travel distance due to an \emph{unrepresentative RMSE} (e.g. not bias-corrected) that overestimates the variability of DEM error.}
\label{fig:hazardmap}
\end{figure}

Figure~\ref{fig:hazardmap}\,(a) and (c) show the probabilistic hazard map for both $\text{USS}_\text{N=500}$ \{RMSE=3.3, d=180\} ensemble and $\text{CSS}_\text{N=500}$ ensemble. It can be seen that the potential hazard area is much larger than the deterministic hazard area for both ensembles. The difference between the deterministic and the ensemble based hazard area is most pronounced in area~1-3 for $\text{USS}_\text{N=500}$\,\{RMSE= 3.3, d=180\} ensemble and in area~3 for $\text{CSS}_\text{N=500}$ ensemble.  Figure~\ref{fig:hazardmap}\,(b) and (d) show boxplots of the apparent friction angle distribution for both ensembles. It is evident that the apparent friction angle of both ensembles varies largely with respect to the apparent friction angle of the deterministic simulation ($16.80^{\circ}$). $\text{CSS}_\text{N=500}$ ensemble-based apparent friction angle (mean $15.39^{\circ}$) is smaller than $\text{USS}_\text{N=500}$\,\{RMSE=3.3, d=180\} ensemble-based apparent friction angle (mean $16.76^{\circ}$). This can be explained as follows.

Due to DEM uncertainty, topographic characteristics represented in DEM realizations vary from that represented in the original DEM. Specifically,
\begin{enumerate}
\item[$\bullet$] topographic details of the deterministic channel tend to be dampened out from DEM realizations. The topographic details include banks of the channel, as well as relatively high elevation area at the end part of the channel that holds back flow material as shown in figure~\ref{fig:deterministic}\,(b);
\item[$\bullet$] topographic roughness tends to increase.
\end{enumerate}

Whether, where, and to what extent the topographic characteristics in DEM realizations would differ from the original DEM depend on
\begin{enumerate}
\item[$\bullet$] variability of DEM error. Intuitively, the larger the variability, the more likely that topographic details of the deterministic channel would be dampened out, and the larger the topographic roughness in DEM realizations;
\item[$\bullet$] topographic details of the original DEM. If subject to the same DEM error, less 'well defined' topographic characteristics in the original DEM are more likely to change in DEM realizations. For example, along the channel of the HK-DTM, the north bank of the channel near area~1 and the south bank of the channel near area~2 are less 'well defined' compared to other parts of channel banks. Flow material could be more easily diverted to area~1 and area~2 where elevations are relatively low and some local 'small channels' exist. Area~3 could also be regarded as less 'well defined' since it is relatively flat and thus is sensitive to DEM uncertainty \cite{Temme2009}. 
\end{enumerate}

The change of each topographic characteristic has corresponding impact on landslide run-out behaviour. Specifically,
\begin{enumerate}
\item[$\bullet$] if banks of the deterministic channel were dampened out in DEM realizations, flow material tends to spread out along channel cross section direction and travel distance would be shorter;
\item[$\bullet$] if the relatively high elevation area that holds back flow material was dampened out, flow material tends to travel further;
\item[$\bullet$] increasing topographic roughness leads to higher simulated momentum losses and shorter travel distance as pointed out by \citet{McDougall2017}.
\end{enumerate}

For $\text{USS}_\text{N=500}$\,\{RMSE=3.3, d=180\} ensemble, the variability of DEM error is relatively large, e.g. around 3.3\,$m$ governed by the not-bias-corrected RMSE based on $\bm{\epsilon}^*_{KK}\{K=180\}$ (see figure~\ref{fig:mean and std grid}\,(c)). In this situation, both the north bank near area~1 and south bank near area~2 as well as the relatively high elevation area at the end part of the channel are possible to be dampened out in HK-DTM realizations. For $\text{CSS}_\text{N=500}$ ensemble, the variability of DEM error is relatively small, e.g. around 1.5\,$m$ governed by the standard deviation ($\sigma$) based on $\bm{\epsilon}^*_{KK}\{K=180\}$ (see figure~\ref{fig:mean and std grid}\,(d)). In this situation, the banks tend to remain 'well defined' while the relatively high elevation area is possible to be dampened out in HK-DTM realizations. Thus, area~1 and area~2 are possibly subject to hazard in $\text{USS}_\text{N=500}$\,\{RMSE=3.3, d=180\} ensemble but less likely in $\text{CSS}_\text{N=500}$ ensemble. As mentioned above, area~3 is a flat area which is sensitive to DEM uncertainty. Furthermore, it locates near the deposition, around which the impact of upstream DEM uncertainty seems to accumulate. Thus, it is highly affected in both ensembles.

The apparent friction angle distribution is determined by a combined effect of change of channel banks, change of the relatively high elevation area at the end part of the channel, and increasing topographic roughness. For $\text{USS}_\text{N=500}$\,\{RMSE=3.3, d=180\} ensemble, deteriorated channel bank representation and increasing topographic roughness make flow material travel shorter distance, e.g. larger apparent friction angle, while deteriorated relatively high elevation area representation allows flow material to travel further, e.g. smaller apparent friction angle. For $\text{CSS}_\text{N=500}$ ensemble, channel banks are likely to remain 'well defined' and the degree of topographic roughness increase is lower due to its relatively small variability of DEM error compared to $\text{USS}_\text{N=500}$\,\{RMSE=3.3, d=180\} ensemble. Thus, flow material in $\text{CSS}_\text{N=500}$ ensemble tends to travel longer distance, e.g. smaller apparent friction angle, compared to $\text{USS}_\text{N=500}$\,\{RMSE=3.3, d=180\} ensemble.

In summary, we can conclude from the probabilistic hazard maps and boxplots of apparent friction angle distribution that: 
\begin{enumerate}
\item[$\bullet$] accounting for DEM uncertainty may significantly increase the potential hazard area; 
\item[$\bullet$] the potential hazard area is highly related to the variability of DEM error and topographic characteristics of the original DEM; 
\item[$\bullet$] USS based on the RMSE only may overestimate the spread of potential hazard area and underestimate travel distance due to a not-bias-corrected RMSE that overestimates the variability of DEM error.
\end{enumerate}
It should be noted that the probabilistic hazard map here is constructed based on maximum height and a pre-defined threshold. In simulation-based hazard assessment practice, one may indicate potential hazard using other indicators, e.g. maximum momentum that reflects the impact pressure, etc. and correspondingly modify the threshold value. In this case, our workflow is easily extendible to account for other indicators.

\subsubsection{Dynamic flow properties}
\label{S:6.2.2}
\begin{figure}[htbp]
\centering
\includegraphics[scale=0.65]{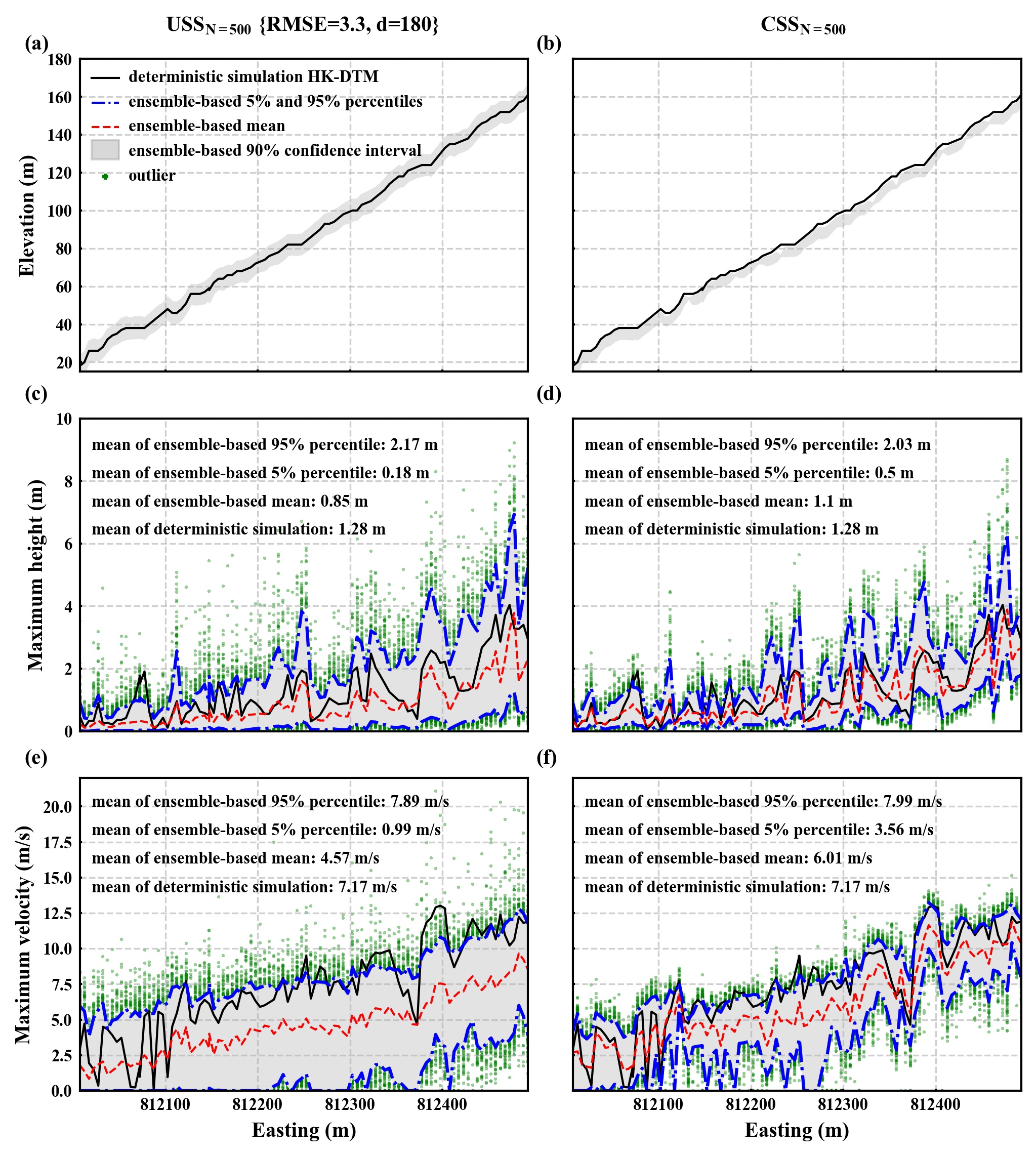}
\caption{Elevation, maximum height, and maximum velocity at locations along the channel bottom (see figure~\ref{fig:deterministic}\,(b)). Left and right columns correspond to $\text{USS}_\text{N=500}$ \{RMSE=3.3, d=180\} ensemble and $\text{CSS}_\text{N=500}$ ensemble respectively. In each subfigure, blue dashed dotted lines represent ensemble-based 5\% and 95\% percentiles of the quantity. The red dashed line represents ensemble-based mean of the quantity. The black line denotes corresponding results of the deterministic simulation. Annotated mean values are average of all the locations. Ensemble-based flow dynamic properties largely vary from deterministic simulation results. The variation range of $\text{USS}_\text{N=500}$ \{RMSE=3.3, d=180\} ensemble is larger while its ensemble-based mean is smaller, compared to counterparts of $\text{CSS}_\text{N=500}$ ensemble. Our main findings are: 1) accounting for DEM uncertainty may significantly affect dynamic flow properties hence any hazard assessment that is based on landslide dynamics; 2) USS based on the RMSE only may overestimate the range of dynamic flow properties and underestimate ensemble-based mean of dynamic flow properties due to an \emph{unrepresentative RMSE} that overestimates the variability of DEM error.}
\label{fig:channel bottom}
\end{figure}

The left column in figure~\ref{fig:channel bottom} shows elevation, maximum height and maximum velocity at locations along the channel bottom based on $\text{USS}_\text{N=500}$ \{RMSE=3.3, d=180\} ensemble. It is evident that both maximum height and maximum velocity at these locations largely vary from that of the deterministic simulation. Specifically, the mean of maximum height (maximum velocity) values at all the locations based on the deterministic simulation is 1.28\,$m$ (7.17\,$m/s$). The mean of ensemble-based 90\% confidence interval of maximum height (maximum velocity) is $[0.18\,m, 2.17\,m]$ ($[0.99\,m/s, 7.89\,m/s]$) (e.g. the range between the mean of ensemble-based 5\% percentile and the mean of ensemble-based 95\% percentile). Another observation is that ensemble-based mean of flow dynamic properties is generally smaller than the mean of flow dynamic properties of the deterministic simulation (e.g. the red dashed line is generally under the black line in both figure~\ref{fig:channel bottom}\,(c) and (e)). The mean of ensemble-based mean of maximum height (maximum velocity) is 0.85\,$m$ (4.57\,$m/s$), around 66\% (64\%) of the mean of the deterministic simulation 1.28\,$m$ (7.17\,$m/s$) (see figure~\ref{fig:channel bottom}\,(c) and (e)).

The right column in figure~\ref{fig:channel bottom} shows corresponding results based on $\text{CSS}_\text{N=500}$ ensemble. Similar trends as in $\text{USS}_\text{N=500}$ \{RMSE=3.3, d=180\} ensemble can also be observed. Namely, both maximum height and velocity at these locations largely vary from that of the deterministic simulation, and ensemble-based mean of flow dynamic properties is generally smaller than deterministic results. Main differences are that the variation range of $\text{CSS}_\text{N=500}$ ensemble-based flow dynamic properties is smaller, and $\text{CSS}_\text{N=500}$ ensemble-based mean of flow dynamic properties is larger compared to $\text{USS}_\text{N=500}$ \{RMSE=3.3, d=180\} ensemble. More specifically, the mean of $\text{CSS}_\text{N=500}$ ensemble-based 90\% confidence interval of maximum height (maximum velocity) is $[0.5\,m, 2.03\,m]$ ($[3.56\,m/s, 7.99\,m/s]$). The mean of $\text{CSS}_\text{N=500}$ ensemble-based mean of maximum height (maximum velocity) is 1.1\,$m$ (6.01\,$m/s$), around 86\% (84\%) of the mean of the deterministic simulation 1.28\,$m$ (7.17 $m/s$) (see figure~\ref{fig:channel bottom}\,(d) and (f)).

\begin{figure}[htbp]
\centering
\includegraphics[scale=0.65]{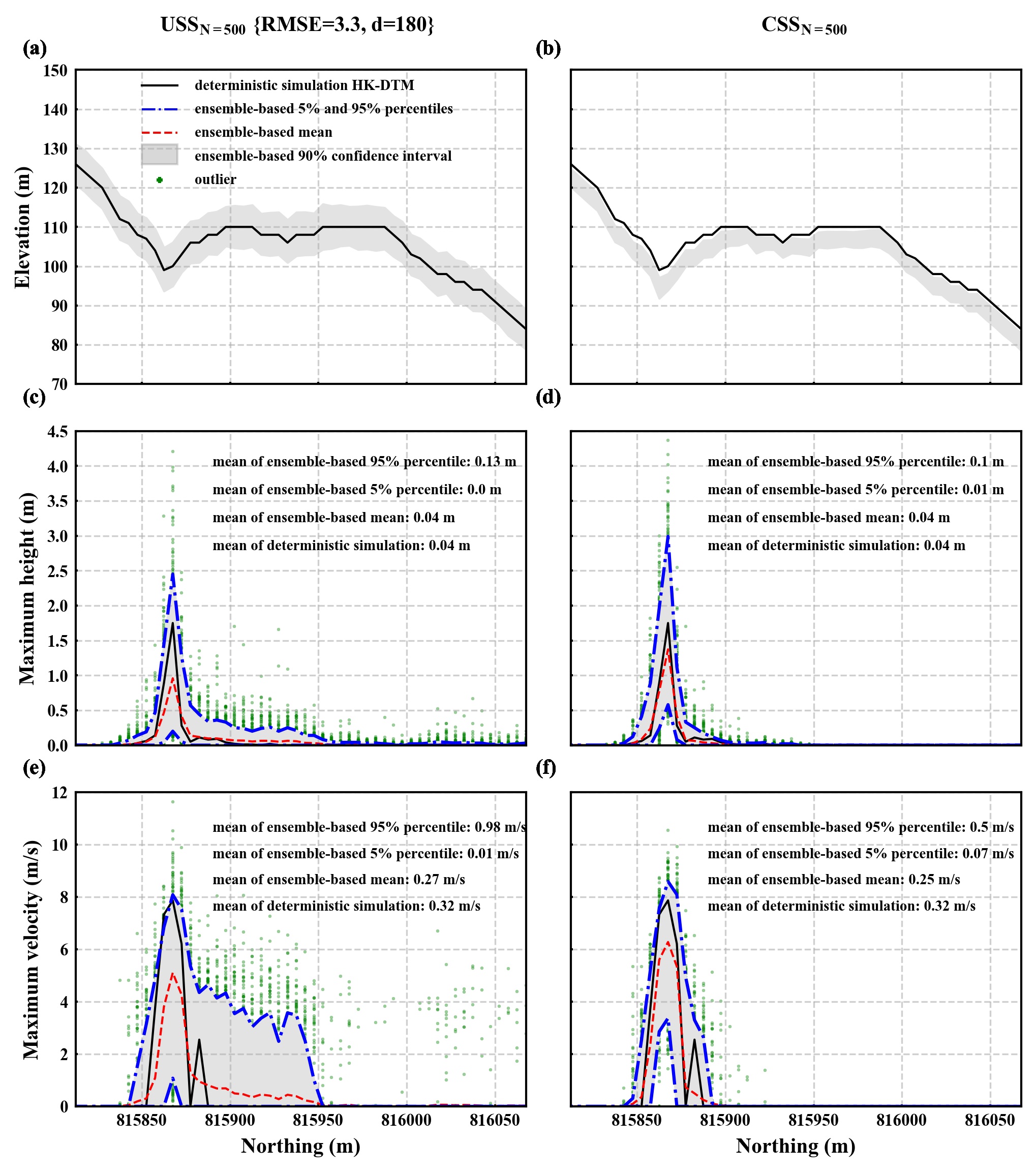}
\caption{Elevation, maximum height, and maximum velocity at locations along the channel cross section (see figure~\ref{fig:deterministic}\,(b)). Left and right columns correspond to $\text{USS}_\text{N=500}$ \{RMSE=3.3, d=180\} ensemble and $\text{CSS}_\text{N=500}$ ensemble respectively. In each subfigure, blue dashed dotted lines represent ensemble-based 5\% and 95\% percentiles of the quantity. The red dashed line represents ensemble-based mean of the quantity. The black line denotes corresponding results of the deterministic simulation. Annotated mean values are average of all the locations. Due to DEM uncertainty, flow material of both ensembles tends to spread out along the channel cross section direction. The ensemble-based mean of flow dynamic properties at the channel bottom location is smaller than flow dynamic properties at the channel bottom location of the deterministic simulation (compare peak value of red dashed line with peak value of black line). The more the flow material spreads out, the smaller the ensemble-based mean of flow dynamic properties at the channel bottom location (compare results of $\text{USS}_\text{N=500}$ \{RMSE=3.3, d=180\} ensemble with that of $\text{CSS}_\text{N=500}$ ensemble.}
\label{fig:cross section}
\end{figure}

The above observations result from similar factors as discussed in section~\ref{S:6.2.1}. Due to DEM uncertainty,
\begin{enumerate}
\item[$\bullet$] ensemble-based flow dynamic properties are likely to vary from that of the deterministic simulation. Larger variability of DEM error is likely to result in more extreme values. As discussed in section~\ref{S:6.2.1}, the variability of DEM error for $\text{USS}_\text{N=500}$ \{RMSE=3.3, d=180\} ensemble is larger than that for $\text{CSS}_\text{N=500}$ ensemble due to \emph{unrepresentative RMSE} issue. Thus the variation range of $\text{USS}_\text{N=500}$ \{RMSE=3.3, d=180\} ensemble-based flow dynamic properties is generally larger than that of $\text{CSS}_\text{N=500}$ ensemble-based flow dynamic properties, e.g. larger mean of ensemble-based 90\% confidence interval (the trend would be more clear if we also consider outliers outside 90\% confidence interval).
\item[$\bullet$] banks of the deterministic channel may be dampened out in DEM realizations. Deteriorated channel bank representation makes flow material more spread out along channel cross section direction. This could lead to smaller ensemble-based mean of flow dynamic properties at channel bottom locations, compared to flow dynamic properties of the deterministic simulation. It can be directly seen in figure~~\ref{fig:cross section}, which displays results of one channel cross section. Also, due to larger variability of DEM error, flow material in $\text{USS}_\text{N=500}$ \{RMSE=3.3, d=180\} ensemble is more spread along channel cross section direction, resulting in smaller ensemble-based mean of flow dynamic properties at channel bottom locations compared to $\text{CSS}_\text{N=500}$ ensemble. This can also be seen in figure~\ref{fig:cross section}.  
\item[$\bullet$] topographic roughness in DEM realizations tends to increase. As pointed out in section~\ref{S:6.2.1}, increasing topographic roughness results in higher simulated momentum losses and thus smaller flow dynamic properties on average. The higher the degree of topographic roughness increase, the higher the simulated momentum losses and the smaller the flow dynamic properties. This also contributes to smaller ensemble-based mean of flow dynamic properties at channel bottom locations, compared to flow dynamic properties of the deterministic simulation, as well as smaller $\text{USS}_\text{N=500}$ \{RMSE=3.3, d=180\} ensemble-based mean of flow dynamic properties at channel bottom locations, compared to $\text{CSS}_\text{N=500}$ ensemble.
\end{enumerate}

Based on the ensembles' dynamic flow properties we can conclude that:
\begin{enumerate}
\item[$\bullet$] accounting for DEM uncertainty may significantly affect dynamic flow properties, e.g. maximum height and maximum velocity, hence any hazard assessment that is based on landslide dynamics; 
\item[$\bullet$] USS based on the RMSE only may overestimate the range of dynamic flow properties and underestimate ensemble-based mean of dynamic flow properties due to an \emph{unrepresentative RMSE} that overestimates the variability of DEM error.
\end{enumerate}

\subsection{Additional ensembles to investigate USS sensitivities in RMSE and d}
\label{S:6.3}

Here, we discuss the \emph{unrepresentative RMSE} and \emph{subjective d} issues as introduced in section~\ref{S:3.1} based on additional six ensembles $\text{USS}_\text{N=500}$ \{RMSE=0.5, 1.5, 2.5, d=180\} and \{RMSE=3.3, d=0, 90, 270\} (refer to section~\ref{S:5.3.2}) as well as the $\text{USS}_\text{N=500}$ \{RMSE=3.3, d=180\} ensemble. Results of the $\text{CSS}_\text{N=500}$ ensemble are used as a reference since $\text{CSS}_\text{N=500}$ incorporated more information on the DEM error. It is thus reasonable to assume that its results reflect the reality better.

\begin{figure}[htb]
\centering
\includegraphics[scale=0.86]{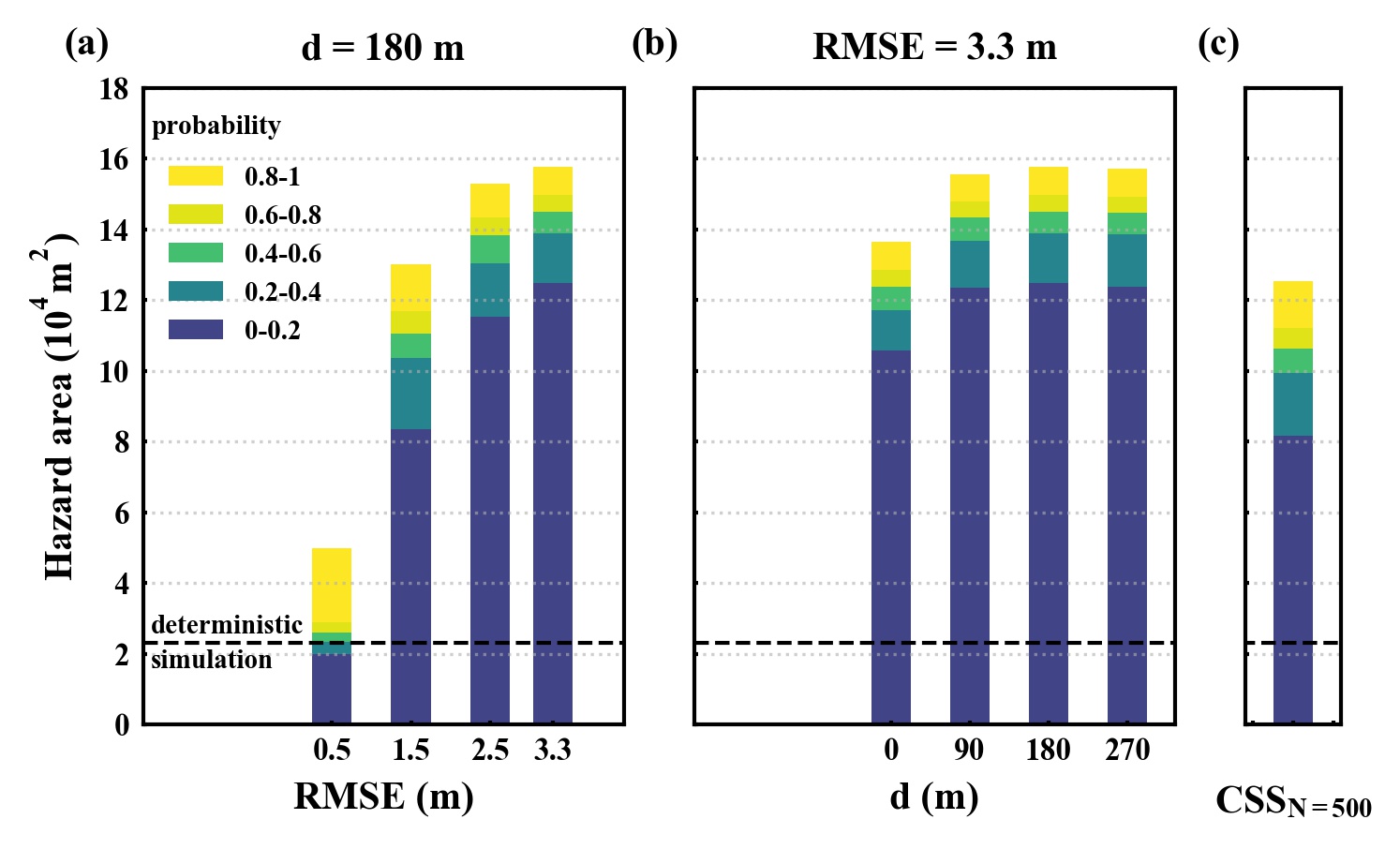}
\caption{Stacked bar plots of the potential hazard area's magnitude based on the probabilistic hazard map for each ensemble (see figure~\ref{fig:hazardmap}\,(a) and (c)). (a) Set of $\text{USS}_\text{N=500}$ \{RMSE=0.5, 1.5, 2.5, 3.3, d=180\} ensembles; (b) set of $\text{USS}_\text{N=500}$ \{RMSE=3.3, d=0, 90, 180, 270\} ensembles; (c) $\text{CSS}_\text{N=500}$ ensemble. Each stacked bar shows the potential hazard area with different hazard probabilities for the corresponding ensemble. The dashed line represents the deterministic hazard area of the deterministic simulation run.}
\label{fig:all ensembles hazardarea}
\end{figure}

\begin{figure}[htb]
\centering
\includegraphics[scale=0.86]{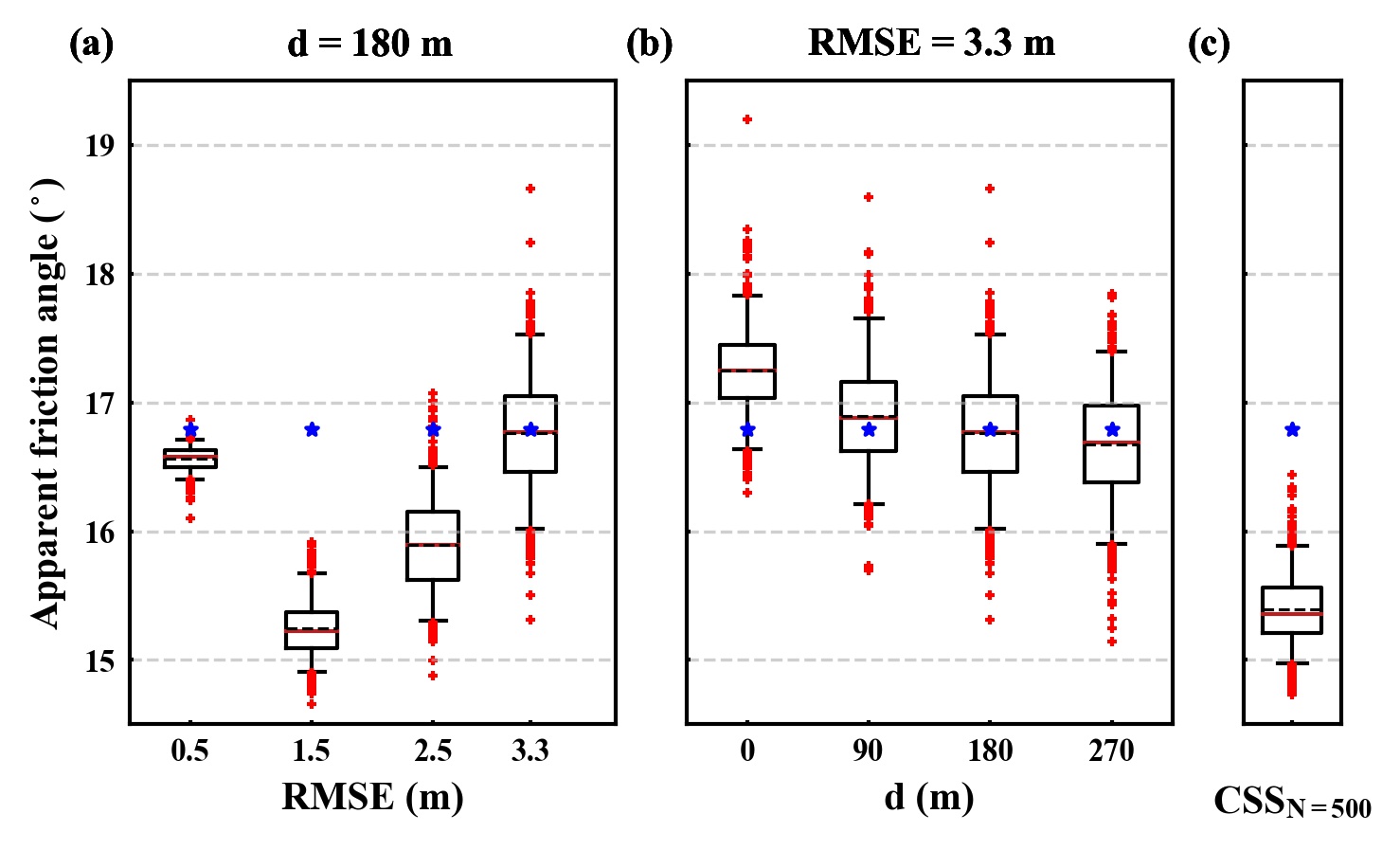}
\caption{Apparent friction angle distribution. (a) Set of $\text{USS}_\text{N=500}$ \{RMSE=0.5, 1.5, 2.5, 3.3, d=180\} ensembles; (b) set of $\text{USS}_\text{N=500}$ \{RMSE=3.3, d=0, 90, 180, 270\} ensembles; (c) $\text{CSS}_\text{N=500}$ ensemble. Each boxplot shows the distribution of apparent friction angle for the corresponding ensemble. The blue star represents the apparent friction angle of the deterministic simulation.}
\label{fig:all ensembles boxplots}
\end{figure}

\begin{figure}[htb]
\centering
\includegraphics[scale=0.86]{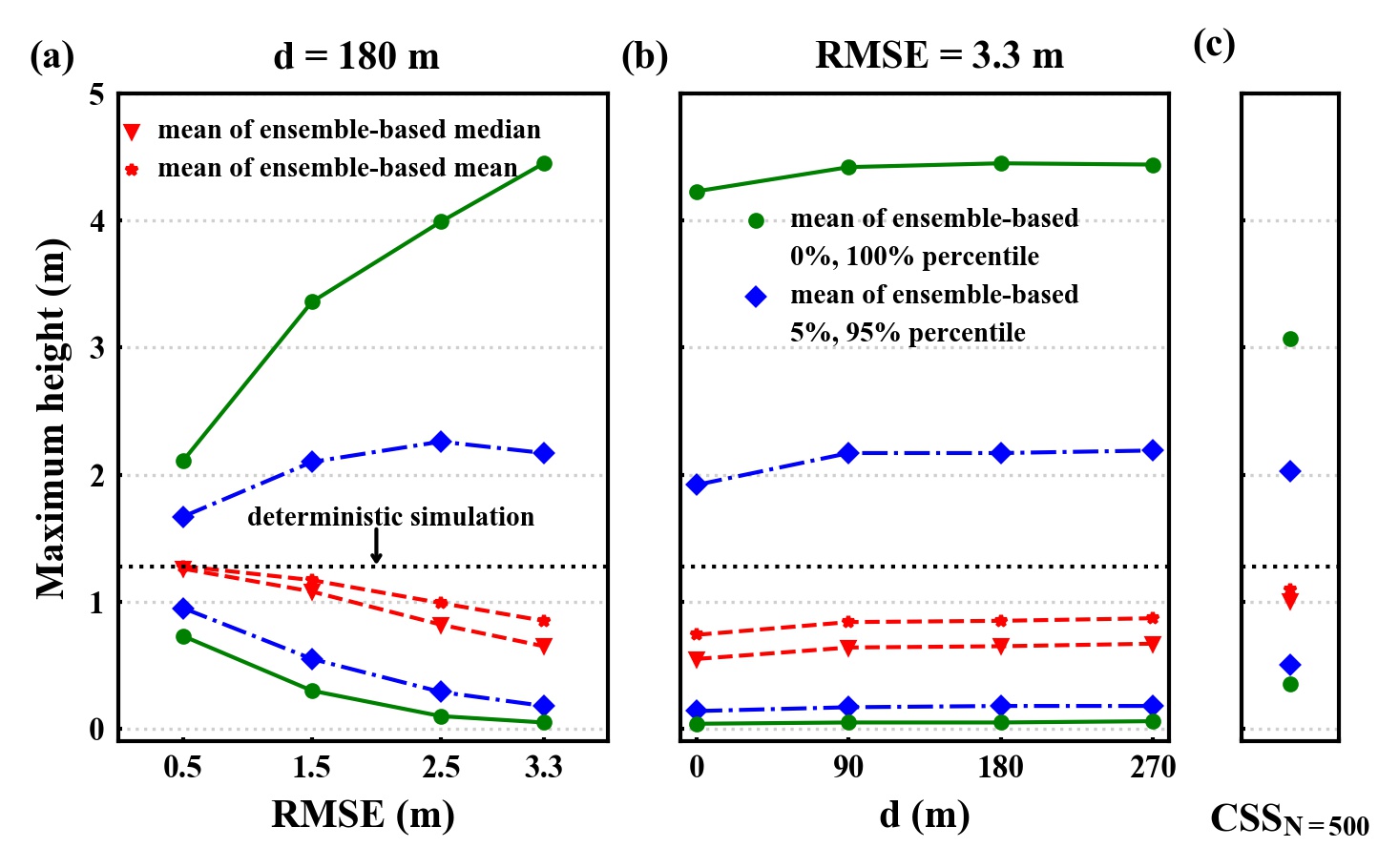}
\caption{Statistics of maximum height at channel bottom locations (see figure~\ref{fig:channel bottom}\,(c) and (d)). (a) Set of $\text{USS}_\text{N=500}$ \{RMSE=0.5, 1.5, 2.5, 3.3, d=180\} ensembles; (b) set of $\text{USS}_\text{N=500}$ \{RMSE=3.3, d=0, 90, 180 , 270\} ensembles; (c) $\text{CSS}_\text{N=500}$ ensemble. The black dotted line shows the average value of maximum height at all the channel bottom locations based on the deterministic simulation.}
\label{fig:all ensembles maxH}
\end{figure}

\begin{figure}[htb]
\centering
\includegraphics[scale=0.86]{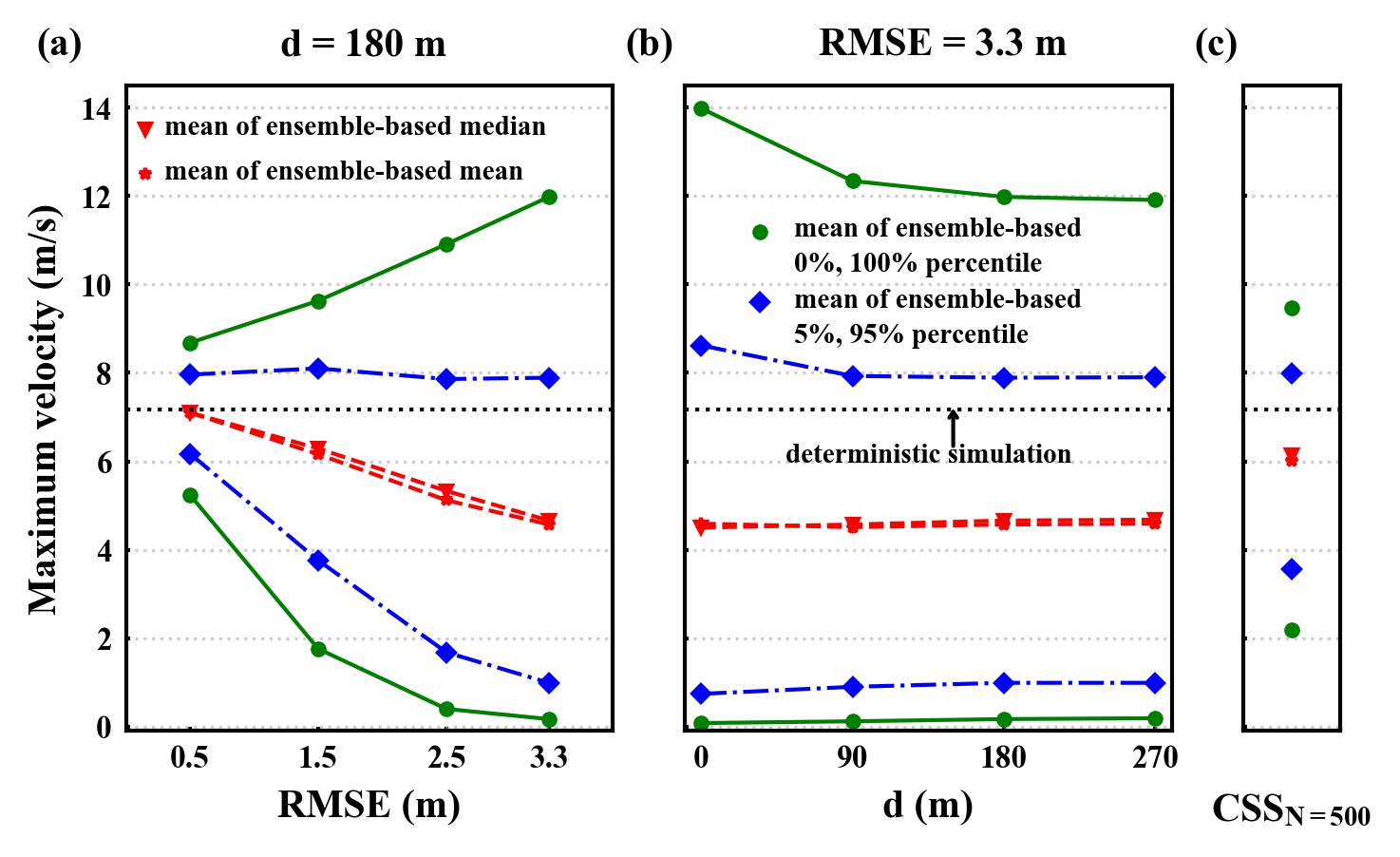}
\caption{Statistics of maximum velocity at channel bottom locations (see figure~\ref{fig:channel bottom}\,(e) and (f)). (a) Set of $\text{USS}_\text{N=500}$ \{RMSE=0.5, 1.5, 2.5, 3.3, d=180\} ensembles; (b) set of $\text{USS}_\text{N=500}$ \{RMSE=3.3, d=0, 90, 180, 270\} ensembles; (c) $\text{CSS}_\text{N=500}$ ensemble. The black dotted line shows the average value of maximum velocity at all the channel bottom locations based on the deterministic simulation.}
\label{fig:all ensembles maxV}
\end{figure}

Figures~\ref{fig:all ensembles hazardarea}-\ref{fig:all ensembles maxV} show the consolidated results of the ensembles. In each figure, subfigure\,(a) corresponds to the set of $\text{USS}_\text{N=500}$ \{RMSE= 0.5, 1.5, 2.5, 3.3, d=180\} ensembles, subfigure\,(b) to the set of $\text{USS}_\text{N=500}$ \{RMSE=3.3, d=0, 90, 180, 270\} ensembles, and subfigure\,(c) to the $\text{CSS}_\text{N=500}$ ensemble. Also, deterministic simulation results are included. For probabilistic hazard maps of the additional ensembles, we refer to the appendix.

Focusing on figures~\ref{fig:all ensembles hazardarea}-\ref{fig:all ensembles maxV}\,(a), it can be seen that with increasing RMSE: 
\begin{enumerate}
\item[$\bullet$] low-probability (0-0.2) hazard area significantly increases and high-probability (0.8-1) hazard area gradually decreases leading to increase of overall potential hazard area if we keep the same threshold value; 
\item[$\bullet$] except for the RMSE=0.5\,m ensemble, the apparent friction angle steadily increases; 
\item[$\bullet$] the range of extreme values of maximum height (maximum velocity) at channel bottom locations increases while average of maximum height (maximum velocity) at channel bottom locations decreases.
\end{enumerate}

For purely RMSE-based USS, the standard deviation of DEM error is assumed to be determined by the RMSE. Hence larger RMSE indicates larger variability of DEM error in DEM realizations. The larger the variability of DEM error, the more likely topographic details of the deterministic channel would be dampened out, and the larger the topographic roughness in DEM realizations. As discussed in section~\ref{S:6.2.1}, this would make flow material more spread out along channel cross section direction (namely larger potential hazard area) and travel shorter distance (namely larger apparent friction angle). As discussed in section ~\ref{S:6.2.2}, larger variability of DEM error is likely to result in more extreme values of flow dynamic properties (namely larger range of extreme values) while spreading of flow material along channel cross section direction and larger topographic roughness lead to smaller ensemble-based mean of flow dynamic properties at channel bottom locations.

As discussed in section~\ref{S:6.2.1}, the apparent friction angle distribution is determined by a combined effect of change of channel banks, change of the relatively high elevation area at the end part of the channel, and increasing topographic roughness. It naturally follows that for a small variability of DEM error (here RMSE=0.5\,$m$), all the changes are less significant in DEM realizations and thus the apparent friction angle of $\text{USS}_\text{N=500}$\{RMSE=0.5, d=180\} closely matches the deterministic simulation result. For an intermediate variability of DEM error (here RMSE=1.5\,$m$), the relatively high elevation area at the end part of the channel is subject to change while channel banks tend to remain 'well defined' in DEM realizations. This leads to longer travel distance of $\text{USS}_\text{N=500}$\{RMSE=1.5, d=180\} ensemble (namely smaller apparent friction angle) in comparison to the deterministic simulation result.

From subfigures~\ref{fig:all ensembles hazardarea}-\ref{fig:all ensembles maxV}\,(b), we find that consistently the results for a USS ensemble of vanishing spatial autocorrelation $\text{USS}_\text{N=500}$\{RMSE=3.3, d=0\} differ significantly from USS ensembles that include spatial autocorrelation, hence $\text{USS}_\text{N=500}$\{RMSE=3.3, d=90, 180, 270\} ensembles. This indicates that whether spatial autocorrelation is considered or not may make a difference but the extent of spatial autocorrelation has less influence on simulation results. As we know spatial autocorrelation to be present in topographic data but often lack information on its exact autocorrelation length, this is actually good news for practical hazard assessment studies. 

Comparing subfigures~\ref{fig:all ensembles hazardarea}-\ref{fig:all ensembles maxV}\,(a) with (c), it can furthermore be seen that the results of the $\text{USS}_\text{N=500}$ \{RMSE=1.5, d=180\} ensemble are quite close to the results of the $\text{CSS}_\text{N=500}$ ensemble. The $\text{USS}_\text{N=500}$ \{RMSE=1.5, d=180\} ensemble is informed with the bias-corrected RMSE (namely the true standard deviation, in our case 1.5\,$m$, see figure~\ref{fig:reference data}\,(b)). It indicates that if a bias-corrected RMSE is given, USS is possible to provide reasonable results considering the extent of spatial autocorrelation has less influence on simulation results.

\clearpage
All in all, we find that: 
\begin{enumerate}
\item[$\bullet$] the results of USS are in general more sensitive to values of the RMSE and less sensitive to values of d; 
\item[$\bullet$] an \emph{unrepresentative RMSE} that overestimates the variability of DEM error may overestimate the potential hazard area and extreme values of dynamic flow properties; 
\item[$\bullet$] whether or not spatial autocorrelation of DEM error is considered can make a difference of ensemble-based simulation results;
\item[$\bullet$] if a bias-corrected RMSE is given, it is possible to obtain reasonable ensemble-based results using USS.
\end{enumerate}

\section{Conclusions}
\label{S:7}
In this paper, we investigated different approaches to study the impact of topographic uncertainty on simulation-based flow-like landslide run-out analyses. Based upon a historic landslide event for which two DEM data sets of different accuracy had been available, we presented a series of computational scenarios. Unconditional and conditional stochastic simulation are conducted to generate DEM realizations, both for the case in which only the RMSE is available, and for the case in which reference data of higher accuracy is available. The computational workflow including the stochastic simulation to generate the DEM realizations and the landslide run-out simulation is implemented as a modular Python-based package. How topographic uncertainty propagates into results of landslide run-out analysis is discussed in detail. In addition, we addressed the two major issues of purely RMSE-based unconditional stochastic simulation, e.g. the fact that not-bias-corrected RMSE overestimates the variability of DEM error (referred to as \emph{unrepresentative RMSE} in our study) and the fact that determining the size of the spatially moving filter in the absence of further information on the spatial DEM error structure is often subjective (referred to as \emph{subjective d} in our study). Our main conclusions are:
\begin{enumerate}
\item[$\bullet$] DEM uncertainty significantly affects simulation-based landslide run-out modeling depending on how well the underlying flow path is represented, which is determined by topographic characteristics of the original DEM and the variability of DEM error. For the same degree of variability of DEM error, the less 'well defined' parts of the flow path in the original DEM are more likely to be affected and thus leads to change of flow behaviour at these parts. Also, an increasing variability of DEM error leads to an increased impact. More specifically, with increasing variability of the DEM error, the potential hazard area and extreme values of dynamic flow properties are likely to increase. This shows the importance of considering topography induced uncertainty for simulation-based landslide hazard assessment rather than simply trusting results of a deterministic simulation if a high accuracy of DEM source is not guaranteed. Also, a preliminary terrain analysis may give some hints on areas that will potential be affected by a topographic uncertainty study. 
\item[$\bullet$] Both unconditional and conditional stochastic simulation methods can be applied to study DEM uncertainty propagation in landslide run-out modeling. Their main difference is that the computationally performant unconditional stochastic simulation can be conducted based on RMSE information only, while the computationally costly conditional stochastic simulation requires the availability of higher accurate reference data and is thus more accurate. The higher accurate reference data provides additional information on the DEM error structure, e.g. its spatial autocorrelation. If the DEM does not contain systematic bias or alternatively a bias-corrected RMSE is provided, the unconditional stochastic simulation yields reasonable results. Otherwise, the assumptions underlying the unconditional stochastic simulation lead to an overestimation of the DEM error variability, which leads to an increased potential impact of DEM uncertainty on the potential hazard area and extreme values of dynamic flow properties. In particular, our study shows that if no higher accurate reference data is available or if computational costs of an conditional stochastic simulation would be too large, the results of a RMSE-based unconditional stochastic simulation can still be used to provide an upper bound of the potential hazard area as well as extreme values of flow dynamic properties for a hazard assessment to take topographic uncertainties into account.
\item[$\bullet$] Results of an unconditional stochastic simulation are in general sensitive to the RMSE value as well as sensitive to the fact whether or not the DEM error's spatial autocorrelation is considered. If the latter is taken into account, results are less sensitive to actual value of the DEM error's maximum autocorrelation length. This indicates that determining a representative RMSE may be more important than finding a correct maximum autocorrelation length, while the DEM error's spatial autocorrelation should not be ignored for simulation-based landslide hazard assessment.
\end{enumerate}

\section*{Acknowledgement}
This work was supported by the China Scholarship Council [grant number: 201706260262].

\appendix
\clearpage
\appendix
\section{Probabilistic hazard maps of all $\text{USS}_\text{N=500}$ ensembles}

\begin{figure}[htbp]
\centering
\includegraphics[scale=0.48]{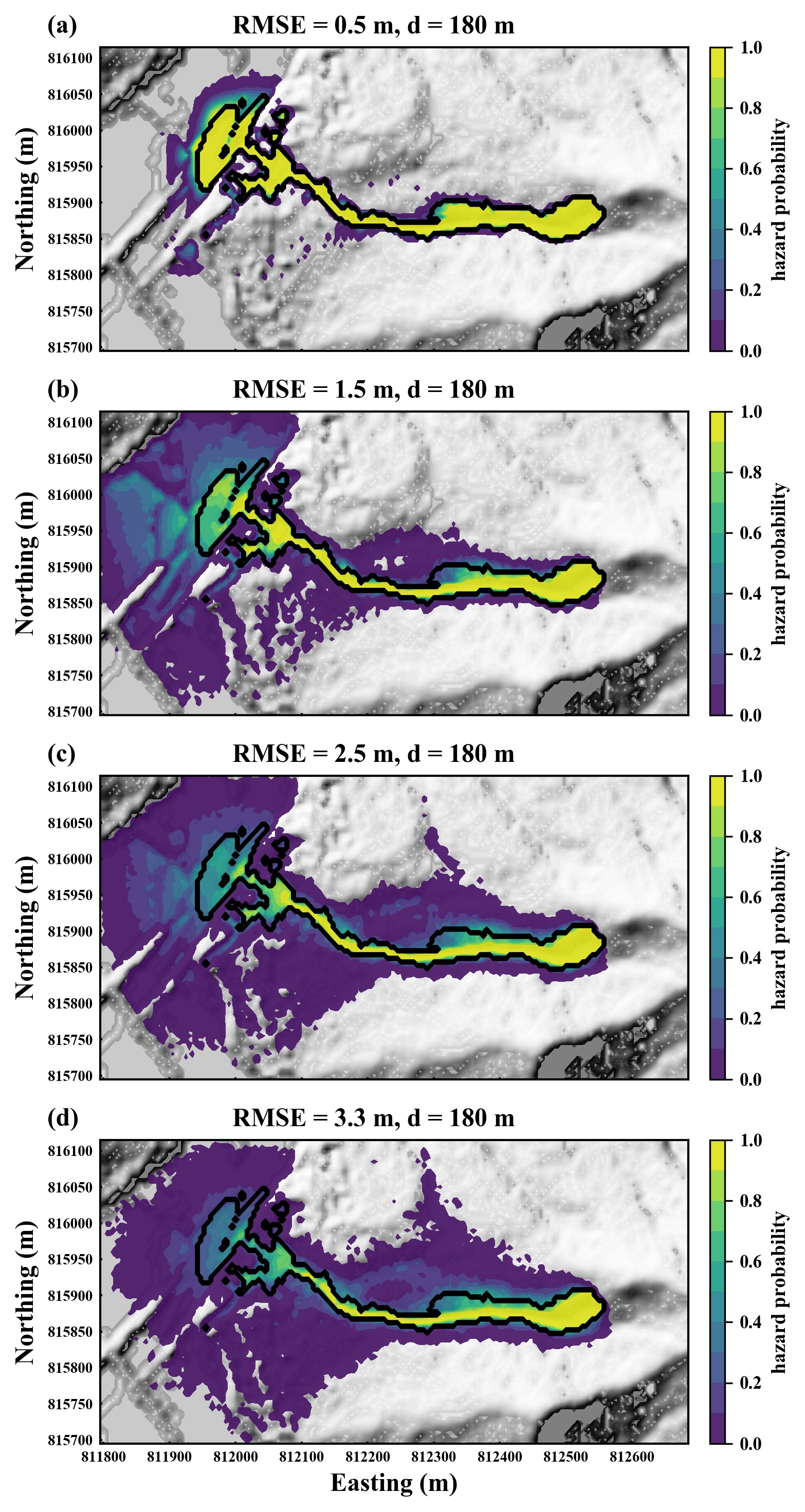}
\caption{Probabilistic hazard map of $\text{USS}_\text{N=500}$ (a) \{RMSE=0.5, d=180\}, (b) \{RMSE=1.5, d=180\}, (c) \{RMSE=2.5, d=180\}, (d) \{RMSE=3.3, d=180\} ensemble. The black outline plotted on the hazard maps represents the deterministic hazard area (see figure~\ref{fig:deterministic}\,(a)).}.
\label{fig:hazardmap RMSE}
\end{figure}

\begin{figure}[htbp]
\centering
\includegraphics[scale=0.48]{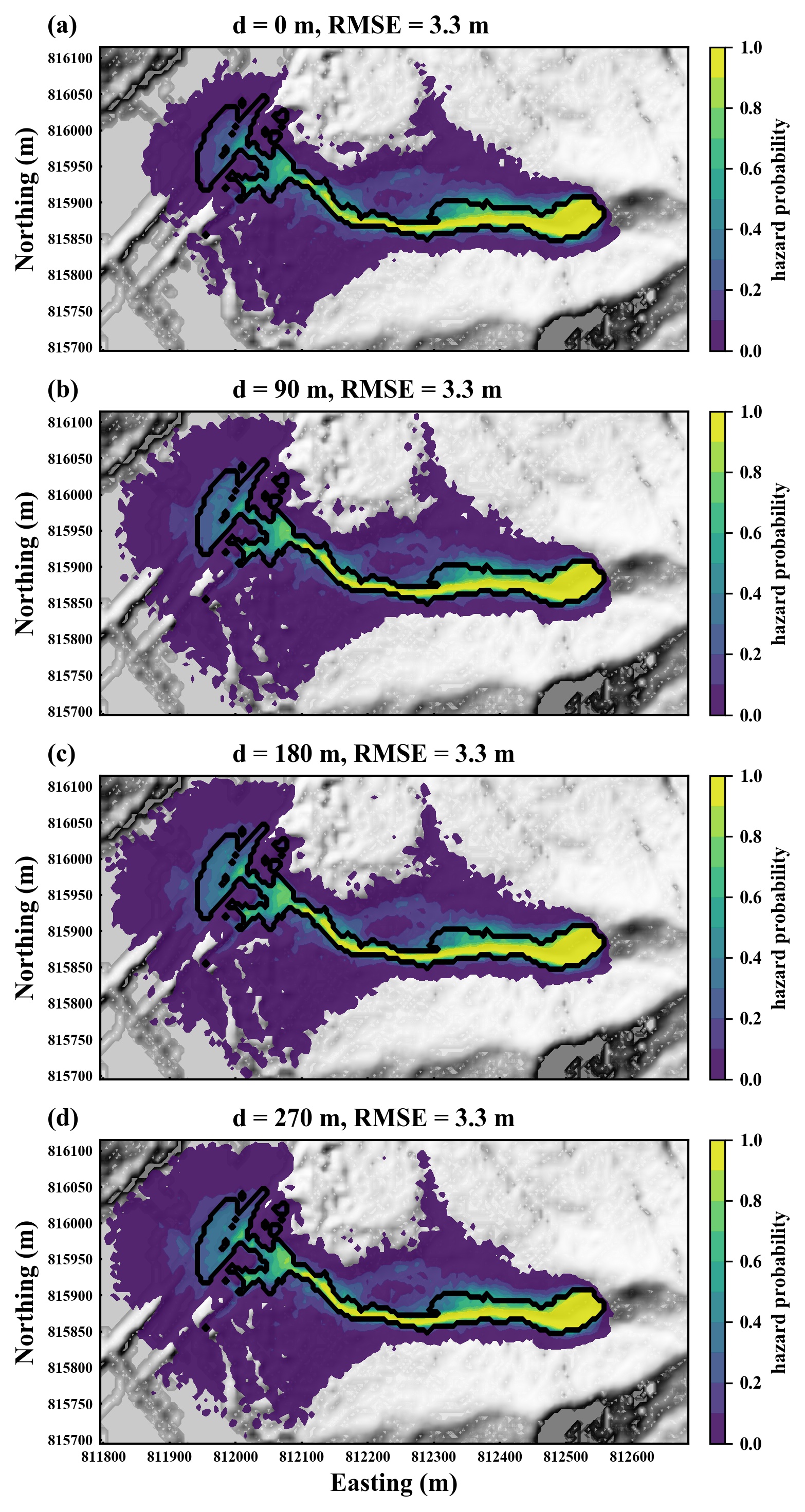}
\caption{Probabilistic hazard map of $\text{USS}_\text{N=500}$ (a) \{RMSE=3.3, d=0\}, (b)\{RMSE=3.3, d=90\}, (c) \{RMSE=3.3, d=180\}, (d) \{RMSE=3.3, d=270\} ensemble. The black outline plotted on the hazard maps represents the deterministic hazard area (see figure~\ref{fig:deterministic}\,(a)).}
\label{fig:hazardmap d}
\end{figure}


\clearpage
\bibliographystyle{elsarticle-template}
\bibliography{topographic_uncertainty}

\begin{thebibliography}{56}
\expandafter\ifx\csname natexlab\endcsname\relax\def\natexlab#1{#1}\fi
\providecommand{\bibinfo}[2]{#2}
\ifx\xfnm\relax \def\xfnm[#1]{\unskip,\space#1}\fi
\bibitem[{Wallemacq et~al.(2018)Wallemacq, UNISDR, and CRED}]{Report2018}
\bibinfo{author}{P.~Wallemacq}, \bibinfo{author}{UNISDR},
  \bibinfo{author}{CRED}, \bibinfo{title}{Economic losses, poverty \& disasters
  1998-2017}, \bibinfo{type}{Technical Report}, \bibinfo{year}{2018}.
\bibitem[{Froude and Petley(2018)}]{Froude2018}
\bibinfo{author}{M.~Froude}, \bibinfo{author}{D.~Petley},
\newblock \bibinfo{title}{Global fatal landslide occurrence from 2004 to 2016},
\newblock \bibinfo{journal}{Natural Hazards and Earth System Sciences}
  \bibinfo{volume}{18} (\bibinfo{year}{2018}) \bibinfo{pages}{2161--2181}.
\bibitem[{Hungr(2009)}]{Hungr2009a}
\bibinfo{author}{O.~Hungr},
\newblock \bibinfo{title}{Numerical modelling of the motion of rapid, flow-like
  landslides for hazard assessment},
\newblock \bibinfo{journal}{KSCE Journal of Civil Engineering}
  \bibinfo{volume}{13} (\bibinfo{year}{2009}) \bibinfo{pages}{281--287}.
\bibitem[{Pastor et~al.(2009)Pastor, Haddad, Sorbino, Cuomo, and
  Drempetic}]{Pastor2009}
\bibinfo{author}{M.~Pastor}, \bibinfo{author}{B.~Haddad},
  \bibinfo{author}{G.~Sorbino}, \bibinfo{author}{S.~Cuomo},
  \bibinfo{author}{V.~Drempetic},
\newblock \bibinfo{title}{A depth-integrated, coupled {SPH} model for flow-like
  landslides and related phenomena},
\newblock \bibinfo{journal}{International Journal for Numerical and Analytical
  Methods in Geomechanics} \bibinfo{volume}{33} (\bibinfo{year}{2009})
  \bibinfo{pages}{143--172}.
\bibitem[{Christen et~al.(2010)Christen, Kowalski, and Bartelt}]{Christen2010}
\bibinfo{author}{M.~Christen}, \bibinfo{author}{J.~Kowalski},
  \bibinfo{author}{P.~Bartelt},
\newblock \bibinfo{title}{{RAMMS}: numerical simulation of dense snow
  avalanches in three-dimensional terrain},
\newblock \bibinfo{journal}{Cold Regions Science and Technology}
  \bibinfo{volume}{63} (\bibinfo{year}{2010}) \bibinfo{pages}{1--14}.
\bibitem[{Xia and Liang(2018)}]{Xia2018}
\bibinfo{author}{X.~Xia}, \bibinfo{author}{Q.~Liang},
\newblock \bibinfo{title}{A new depth-averaged model for flow-like landslides
  over complex terrains with curvatures and steep slopes},
\newblock \bibinfo{journal}{Engineering Geology} \bibinfo{volume}{234}
  (\bibinfo{year}{2018}) \bibinfo{pages}{174--191}.
\bibitem[{Pitman et~al.(2003)Pitman, Nichita, Patra, Bauer, Sheridan, and
  Bursik}]{Pitman2003}
\bibinfo{author}{E.~Pitman}, \bibinfo{author}{C.~Nichita},
  \bibinfo{author}{A.~Patra}, \bibinfo{author}{A.~Bauer},
  \bibinfo{author}{M.~Sheridan}, \bibinfo{author}{M.~Bursik},
\newblock \bibinfo{title}{Computing granular avalanches and landslides},
\newblock \bibinfo{journal}{Physics of Fluids} \bibinfo{volume}{15}
  (\bibinfo{year}{2003}) \bibinfo{pages}{3638--3646}.
\bibitem[{Mast et~al.(2014)Mast, Arduino, Miller, and
  Mackenzie-Helnwein}]{Mast2014}
\bibinfo{author}{C.~Mast}, \bibinfo{author}{P.~Arduino},
  \bibinfo{author}{G.~Miller}, \bibinfo{author}{P.~Mackenzie-Helnwein},
\newblock \bibinfo{title}{Avalanche and landslide simulation using the material
  point method: flow dynamics and force interaction with structures},
\newblock \bibinfo{journal}{Computational Geosciences} \bibinfo{volume}{18}
  (\bibinfo{year}{2014}) \bibinfo{pages}{817--830}.
\bibitem[{Teufelsbauer et~al.(2011)Teufelsbauer, Wang, Pudasaini, Borja, and
  Wu}]{Teufelsbauer2011}
\bibinfo{author}{H.~Teufelsbauer}, \bibinfo{author}{Y.~Wang},
  \bibinfo{author}{S.~Pudasaini}, \bibinfo{author}{R.~Borja},
  \bibinfo{author}{W.~Wu},
\newblock \bibinfo{title}{{DEM} simulation of impact force exerted by granular
  flow on rigid structures},
\newblock \bibinfo{journal}{Acta Geotechnica} \bibinfo{volume}{6}
  (\bibinfo{year}{2011}) \bibinfo{pages}{119--133}.
\bibitem[{McDougall(2017)}]{McDougall2017}
\bibinfo{author}{S.~McDougall},
\newblock \bibinfo{title}{2014 {Canadian} {Geotechnical} {Colloquium}:
  landslide runout analysis -- current practice and challenges},
\newblock \bibinfo{journal}{Canadian Geotechnical Journal} \bibinfo{volume}{54}
  (\bibinfo{year}{2017}) \bibinfo{pages}{605--620}.
\bibitem[{Pastor et~al.(2018)Pastor, Soga, McDougall, and Kwan}]{Pastor2018}
\bibinfo{author}{M.~Pastor}, \bibinfo{author}{K.~Soga},
  \bibinfo{author}{S.~McDougall}, \bibinfo{author}{J.~Kwan},
\newblock \bibinfo{title}{Review of benchmarking exercise on landslide runout
  analysis 2018},
\newblock in: \bibinfo{editor}{K.~Ho}, \bibinfo{editor}{A.~Leung},
  \bibinfo{editor}{J.~Kwan}, \bibinfo{editor}{R.~Koo}, \bibinfo{editor}{R.~Law}
  (Eds.), \bibinfo{booktitle}{Proceedings of the {Second} {JTC1} {Workshop} on
  Triggering and Propagation of Rapid Flow-like Landslides},
  \bibinfo{publisher}{The Hong Kong Geotechnical Society},
  \bibinfo{year}{2018}, pp. \bibinfo{pages}{281--323}.
\bibitem[{Miller and Laflamme(1958)}]{Miller1958}
\bibinfo{author}{C.~Miller}, \bibinfo{author}{R.~Laflamme},
\newblock \bibinfo{title}{The digital terrain model -- theory \& application},
\newblock \bibinfo{journal}{Photogrammetric Engineering} \bibinfo{volume}{24}
  (\bibinfo{year}{1958}) \bibinfo{pages}{433--442}.
\bibitem[{Wilson(2012)}]{Wilson2012}
\bibinfo{author}{J.~Wilson},
\newblock \bibinfo{title}{Digital terrain modeling},
\newblock \bibinfo{journal}{Geomorphology} \bibinfo{volume}{137}
  (\bibinfo{year}{2012}) \bibinfo{pages}{107--121}.
\bibitem[{Wechsler(2007)}]{Wechsler2007}
\bibinfo{author}{S.~Wechsler},
\newblock \bibinfo{title}{Uncertainties associated with digital elevation
  models for hydrologic applications: a review},
\newblock \bibinfo{journal}{Hydrology and Earth system Sciences}
  \bibinfo{volume}{11} (\bibinfo{year}{2007}) \bibinfo{pages}{1481--1500}.
\bibitem[{Rodriguez et~al.(2006)Rodriguez, Morris, and Belz}]{Rodriguez2006}
\bibinfo{author}{E.~Rodriguez}, \bibinfo{author}{C.~Morris},
  \bibinfo{author}{J.~Belz},
\newblock \bibinfo{title}{A global assessment of the {SRTM} performance},
\newblock \bibinfo{journal}{Photogrammetric Engineering \& Remote Sensing}
  \bibinfo{volume}{72} (\bibinfo{year}{2006}) \bibinfo{pages}{249--260}.
\bibitem[{Courty et~al.(2019)Courty, Soriano-Monzalvo, and
  Pedrozo-Acuna}]{Courty2019}
\bibinfo{author}{L.~Courty}, \bibinfo{author}{J.~Soriano-Monzalvo},
  \bibinfo{author}{A.~Pedrozo-Acuna},
\newblock \bibinfo{title}{Evaluation of open-access global digital elevation
  models ({AW3D30}, {SRTM}, and {ASTER}) for flood modelling purposes},
\newblock \bibinfo{journal}{Journal of Flood Risk Management}
  \bibinfo{volume}{12} (\bibinfo{year}{2019}) \bibinfo{pages}{e12550}.
\bibitem[{Wessel et~al.(2018)Wessel, Huber, Wohlfart, Marschalk, Kosmann, and
  Roth}]{Wessel2018}
\bibinfo{author}{B.~Wessel}, \bibinfo{author}{M.~Huber},
  \bibinfo{author}{C.~Wohlfart}, \bibinfo{author}{U.~Marschalk},
  \bibinfo{author}{D.~Kosmann}, \bibinfo{author}{A.~Roth},
\newblock \bibinfo{title}{Accuracy assessment of the global {TanDEM-X} digital
  elevation model with {GPS} data},
\newblock \bibinfo{journal}{ISPRS Journal of Photogrammetry and Remote Sensing}
  \bibinfo{volume}{139} (\bibinfo{year}{2018}) \bibinfo{pages}{171--182}.
\bibitem[{Pakoksung and Takagi(2016)}]{Pakoksung2016}
\bibinfo{author}{K.~Pakoksung}, \bibinfo{author}{M.~Takagi},
\newblock \bibinfo{title}{Digital elevation models on accuracy validation and
  bias correction in vertical},
\newblock \bibinfo{journal}{Modeling Earth Systems and Environment}
  \bibinfo{volume}{2} (\bibinfo{year}{2016}) \bibinfo{pages}{11}.
\bibitem[{Hawker et~al.(2018)Hawker, Bates, Neal, and Rougier}]{Hawker2018}
\bibinfo{author}{L.~Hawker}, \bibinfo{author}{P.~Bates},
  \bibinfo{author}{J.~Neal}, \bibinfo{author}{J.~Rougier},
\newblock \bibinfo{title}{Perspectives on digital elevation model ({DEM})
  simulation for flood modeling in the absence of a high-accuracy open assess
  global {DEM}},
\newblock \bibinfo{journal}{Frontiers in Earth Science} \bibinfo{volume}{6}
  (\bibinfo{year}{2018}) \bibinfo{pages}{233}.
\bibitem[{Krishnan et~al.(2011)Krishnan, Crosby, Nandigam, Phan, Cowart, Baru,
  and Arrowsmith}]{Krishnan2011}
\bibinfo{author}{S.~Krishnan}, \bibinfo{author}{C.~Crosby},
  \bibinfo{author}{V.~Nandigam}, \bibinfo{author}{M.~Phan},
  \bibinfo{author}{C.~Cowart}, \bibinfo{author}{C.~Baru},
  \bibinfo{author}{R.~Arrowsmith},
\newblock \bibinfo{title}{{OpenTopography}: a services oriented architecture
  for community access to {LiDAR} topography},
\newblock in: \bibinfo{booktitle}{ACM International Conference Proceeding
  Series}, p.~\bibinfo{pages}{7}.
\bibitem[{Berry et~al.(2007)Berry, Garlick, and RG}]{Berry2007}
\bibinfo{author}{P.~Berry}, \bibinfo{author}{J.~Garlick},
  \bibinfo{author}{S.~RG},
\newblock \bibinfo{title}{Near-global validation of the {SRTM} {DEM} using
  satellite radar altimetry},
\newblock \bibinfo{journal}{Remote Sensing of Environment}
  \bibinfo{volume}{106} (\bibinfo{year}{2007}) \bibinfo{pages}{17--27}.
\bibitem[{Mouratidis and Ampatzidis(2019)}]{Mouratidis2019}
\bibinfo{author}{A.~Mouratidis}, \bibinfo{author}{D.~Ampatzidis},
\newblock \bibinfo{title}{European digital elevation model validation against
  extensive global navigation satellite systems data and comparison with {SRTM
  DEM and ASTER GDEM in central Macedonia (Greece)}},
\newblock \bibinfo{journal}{International Journal of Geo-Information}
  \bibinfo{volume}{8} (\bibinfo{year}{2019}) \bibinfo{pages}{108}.
\bibitem[{Hofton et~al.(2006)Hofton, Dubayah, Blair, and Rabine}]{Hofton2006}
\bibinfo{author}{M.~Hofton}, \bibinfo{author}{R.~Dubayah},
  \bibinfo{author}{J.~Blair}, \bibinfo{author}{D.~Rabine},
\newblock \bibinfo{title}{Validation of {SRTM} elevations over vegetated and
  non-vegetated terrain using medium footprint {LiDAR}},
\newblock \bibinfo{journal}{Photogrammetric Engineering \& Remote Sensing}
  \bibinfo{volume}{72} (\bibinfo{year}{2006}) \bibinfo{pages}{279--285}.
\bibitem[{Bolkas et~al.(2016)Bolkas, Fotopoulos, Braun, and
  Tziavos}]{Bolkas2016}
\bibinfo{author}{D.~Bolkas}, \bibinfo{author}{G.~Fotopoulos},
  \bibinfo{author}{A.~Braun}, \bibinfo{author}{I.~Tziavos},
\newblock \bibinfo{title}{Assessing digital elevation model uncertainty using
  {GPS} survey data},
\newblock \bibinfo{journal}{Journal of Surveying Engineering}
  \bibinfo{volume}{142} (\bibinfo{year}{2016}) \bibinfo{pages}{04016001}.
\bibitem[{Patel et~al.(2016)Patel, Katiyar, and Prasad}]{Patel2016}
\bibinfo{author}{A.~Patel}, \bibinfo{author}{S.~Katiyar},
  \bibinfo{author}{V.~Prasad},
\newblock \bibinfo{title}{Performances evaluation of different open source
  {DEM} using differential global positioning system ({DGPS})},
\newblock \bibinfo{journal}{The Egyptian Journal of Remote Sensing and Space
  Science} \bibinfo{volume}{19} (\bibinfo{year}{2016}) \bibinfo{pages}{7--16}.
\bibitem[{Elkhrachy(2018)}]{Elkhrachy2018}
\bibinfo{author}{I.~Elkhrachy},
\newblock \bibinfo{title}{Vertical accuracy assessment for {SRTM and ASTER}
  digital elevation models: a case study of {Najran city, Saudi Arabia}},
\newblock \bibinfo{journal}{Ain Shams Engineering Journal} \bibinfo{volume}{9}
  (\bibinfo{year}{2018}) \bibinfo{pages}{1807--1817}.
\bibitem[{Oksanen(2003)}]{Oksanen2003}
\bibinfo{author}{J.~Oksanen},
\newblock \bibinfo{title}{Tracing the gross errors of {DEM}-visualization
  techniques for preliminary quality analysis},
\newblock in: \bibinfo{booktitle}{Proceedings of the 21st International
  Cartographic Conference 'Cartographic Renaissance'}, pp.
  \bibinfo{pages}{2410--2416}.
\bibitem[{Hengl et~al.(2004)Hengl, Gruber, and Shrestha}]{Hengl2004}
\bibinfo{author}{T.~Hengl}, \bibinfo{author}{S.~Gruber},
  \bibinfo{author}{D.~Shrestha},
\newblock \bibinfo{title}{Reduction of errors in digital terrain parameters
  used in soil-landscape modeling},
\newblock \bibinfo{journal}{International Journal of Applied Earth Observation
  and Geoinformation} \bibinfo{volume}{5} (\bibinfo{year}{2004})
  \bibinfo{pages}{97--112}.
\bibitem[{Fisher and Tate(2006)}]{Fisher2006}
\bibinfo{author}{P.~Fisher}, \bibinfo{author}{N.~Tate},
\newblock \bibinfo{title}{Causes and consequences of error in digital elevation
  models},
\newblock \bibinfo{journal}{Progress in Physical Geography: Earth and
  Environment} \bibinfo{volume}{30} (\bibinfo{year}{2006})
  \bibinfo{pages}{467--489}.
\bibitem[{Gonga-Saholiariliva et~al.(2011)Gonga-Saholiariliva, Gunnell, Petit,
  and Mering}]{Gonga-Saholiariliva2011}
\bibinfo{author}{N.~Gonga-Saholiariliva}, \bibinfo{author}{Y.~Gunnell},
  \bibinfo{author}{C.~Petit}, \bibinfo{author}{C.~Mering},
\newblock \bibinfo{title}{Techniques for quantifying the accuracy of gridded
  elevation models and for mapping uncertainty in digital terrain analysis},
\newblock \bibinfo{journal}{Progress in Physical Geography: Earth and
  Environment} \bibinfo{volume}{35} (\bibinfo{year}{2011})
  \bibinfo{pages}{739--764}.
\bibitem[{Holmes et~al.(2000)Holmes, Chadwick, and Kyriakidis}]{Holmes2000}
\bibinfo{author}{K.~Holmes}, \bibinfo{author}{O.~Chadwick},
  \bibinfo{author}{P.~Kyriakidis},
\newblock \bibinfo{title}{Error in a {USGS} 30-meter digital elevation model
  and its impact on terrain modeling},
\newblock \bibinfo{journal}{Journal of Hydrology} \bibinfo{volume}{233}
  (\bibinfo{year}{2000}) \bibinfo{pages}{154--173}.
\bibitem[{Raaflaub and MJ(2006)}]{Raaflaub2006}
\bibinfo{author}{L.~Raaflaub}, \bibinfo{author}{C.~MJ},
\newblock \bibinfo{title}{The effect of error in gridded digital elevation
  models on the estimation of topographic parameters},
\newblock \bibinfo{journal}{Environmental Modelling \& Software}
  \bibinfo{volume}{21} (\bibinfo{year}{2006}) \bibinfo{pages}{710--732}.
\bibitem[{Moawad and EI~Aziz(2018)}]{Moawad2018}
\bibinfo{author}{M.~Moawad}, \bibinfo{author}{A.~EI~Aziz},
\newblock \bibinfo{title}{Assessment of remotely sensed digital elevation
  models ({DEMs}) compared with {DGPS} elevation data and its influence on
  topographic attributes},
\newblock \bibinfo{journal}{Advances in Remote Sensing} \bibinfo{volume}{7}
  (\bibinfo{year}{2018}) \bibinfo{pages}{144--162}.
\bibitem[{Watson et~al.(2015)Watson, Carrivick, and Quincey}]{Watson2015}
\bibinfo{author}{C.~Watson}, \bibinfo{author}{J.~Carrivick},
  \bibinfo{author}{D.~Quincey},
\newblock \bibinfo{title}{An improved method to represent {DEM} uncertainty in
  glacial lake outburst flood propagation using stochastic simulations},
\newblock \bibinfo{journal}{Journal of Hydrology} \bibinfo{volume}{529}
  (\bibinfo{year}{2015}) \bibinfo{pages}{1373--1389}.
\bibitem[{Kiczko and Miroslaw-Swiatek(2018)}]{Kiczko2018}
\bibinfo{author}{A.~Kiczko}, \bibinfo{author}{D.~Miroslaw-Swiatek},
\newblock \bibinfo{title}{Impact of uncertainty of floodplain digital terrain
  model on {1D} hydrodynamic flow calculation},
\newblock \bibinfo{journal}{Water} \bibinfo{volume}{10} (\bibinfo{year}{2018})
  \bibinfo{pages}{1308}.
\bibitem[{Aziz et~al.(2012)Aziz, Steward, Kaleita, and Karkee}]{Aziz2012}
\bibinfo{author}{S.~Aziz}, \bibinfo{author}{B.~Steward},
  \bibinfo{author}{A.~Kaleita}, \bibinfo{author}{M.~Karkee},
\newblock \bibinfo{title}{Assessing the effects of {DEM} uncertainty on erosion
  rate estimation in an agricultural field},
\newblock \bibinfo{journal}{Transactions of the ASABE} \bibinfo{volume}{55}
  (\bibinfo{year}{2012}) \bibinfo{pages}{785--798}.
\bibitem[{Qin et~al.(2013)Qin, Bao, Zhu, Wang, and Hu}]{Qin2013}
\bibinfo{author}{C.~Qin}, \bibinfo{author}{L.~Bao}, \bibinfo{author}{A.~Zhu},
  \bibinfo{author}{R.~Wang}, \bibinfo{author}{X.~Hu},
\newblock \bibinfo{title}{Uncertainty due to {DEM} error in landslide
  susceptibility mapping},
\newblock \bibinfo{journal}{International Journal of Geographical Information
  Science} \bibinfo{volume}{27} (\bibinfo{year}{2013})
  \bibinfo{pages}{1364--1380}.
\bibitem[{Stefanescu et~al.(2012)Stefanescu, Bursik, Cordoba, Dalbey, Jones,
  Patra, Pieri, Pitman, and Sheridan}]{Stefanescu2012}
\bibinfo{author}{E.~Stefanescu}, \bibinfo{author}{M.~Bursik},
  \bibinfo{author}{G.~Cordoba}, \bibinfo{author}{K.~Dalbey},
  \bibinfo{author}{M.~Jones}, \bibinfo{author}{A.~Patra},
  \bibinfo{author}{D.~Pieri}, \bibinfo{author}{E.~Pitman},
  \bibinfo{author}{M.~Sheridan},
\newblock \bibinfo{title}{Digital elevation model uncertainty and hazard
  analysis using a geophysical flow model},
\newblock \bibinfo{journal}{Proceedings of the Royal Society A}
  \bibinfo{volume}{468} (\bibinfo{year}{2012}) \bibinfo{pages}{1543--1563}.
\bibitem[{Naef et~al.(2006)Naef, Rickenmann, Rutschmann, and
  McArdell}]{Naef2006}
\bibinfo{author}{D.~Naef}, \bibinfo{author}{D.~Rickenmann},
  \bibinfo{author}{P.~Rutschmann}, \bibinfo{author}{B.~McArdell},
\newblock \bibinfo{title}{Comparison of flow resistance relations for debris
  flows using a one-dimensional finite element simulation model},
\newblock \bibinfo{journal}{Natural Hazards and Earth System Sciences}
  \bibinfo{volume}{6} (\bibinfo{year}{2006}) \bibinfo{pages}{155--165}.
\bibitem[{Hungr and McDougall(2009)}]{Hungr2009b}
\bibinfo{author}{O.~Hungr}, \bibinfo{author}{S.~McDougall},
\newblock \bibinfo{title}{Two numerical models for landslide dynamic analysis},
\newblock \bibinfo{journal}{Computers \& Geosciences} \bibinfo{volume}{35}
  (\bibinfo{year}{2009}) \bibinfo{pages}{978--992}.
\bibitem[{Bartelt et~al.(1999)Bartelt, Salm, and Gruber}]{Bartelt1999}
\bibinfo{author}{P.~Bartelt}, \bibinfo{author}{B.~Salm},
  \bibinfo{author}{U.~Gruber},
\newblock \bibinfo{title}{Calculating dense-snow avalanche runout using a
  voellmy-fluid model with active/passive longitudinal straining},
\newblock \bibinfo{journal}{Journal of Glaciology} \bibinfo{volume}{45}
  (\bibinfo{year}{1999}) \bibinfo{pages}{242--254}.
\bibitem[{Hungr et~al.(2005)Hungr, Corominas, and Eberhardt}]{Hungr2005}
\bibinfo{author}{O.~Hungr}, \bibinfo{author}{J.~Corominas},
  \bibinfo{author}{E.~Eberhardt},
\newblock \bibinfo{title}{Estimating landslide motion mechanism, travel
  distance and velocity},
\newblock in: \bibinfo{editor}{O.~Hungr}, \bibinfo{editor}{R.~Fell},
  \bibinfo{editor}{R.~Couture}, \bibinfo{editor}{E.~Eberhardt} (Eds.),
  \bibinfo{booktitle}{Landslide risk management}, pp. \bibinfo{pages}{99--128}.
\bibitem[{Salm(1993)}]{Salm1993}
\bibinfo{author}{B.~Salm},
\newblock \bibinfo{title}{Flow, flow transition and runout distances of flowing
  avalanches},
\newblock \bibinfo{journal}{Annals of Glaciology} \bibinfo{volume}{18}
  (\bibinfo{year}{1993}) \bibinfo{pages}{221--226}.
\bibitem[{Frank et~al.(2015)Frank, McArdell, Huggel, and Vieli}]{Frank2015}
\bibinfo{author}{F.~Frank}, \bibinfo{author}{B.~McArdell},
  \bibinfo{author}{C.~Huggel}, \bibinfo{author}{A.~Vieli},
\newblock \bibinfo{title}{The importance of entrainment and bulking on debris
  flow runout modeling: examples from the {Swiss Alps}},
\newblock \bibinfo{journal}{Natural Hazards and Earth System Sciences}
  \bibinfo{volume}{15} (\bibinfo{year}{2015}) \bibinfo{pages}{2569--2583}.
\bibitem[{Hussin et~al.(2012)Hussin, Luna, van Westen, Christen, Malet, and van
  Asch}]{Hussin2012}
\bibinfo{author}{H.~Hussin}, \bibinfo{author}{B.~Luna}, \bibinfo{author}{C.~van
  Westen}, \bibinfo{author}{M.~Christen}, \bibinfo{author}{J.~Malet},
  \bibinfo{author}{T.~van Asch},
\newblock \bibinfo{title}{Parameterization of a numerical {2-D} debris flow
  model with entrainment: a case study of the {Faucon catchment, Southern
  French Alps}},
\newblock \bibinfo{journal}{Natural Hazards and Earth System Sciences}
  \bibinfo{volume}{12} (\bibinfo{year}{2012}) \bibinfo{pages}{3075--3090}.
\bibitem[{Kumar et~al.(2019)Kumar, Gupta, Jamir, and Chattoraj}]{Kumar2019}
\bibinfo{author}{V.~Kumar}, \bibinfo{author}{V.~Gupta},
  \bibinfo{author}{I.~Jamir}, \bibinfo{author}{S.~Chattoraj},
\newblock \bibinfo{title}{Evaluation of potential landslide damming: Case study
  of {Urni landslide, Kinnaur, Satluj valley, India}},
\newblock \bibinfo{journal}{Geoscience Frontiers} \bibinfo{volume}{10}
  (\bibinfo{year}{2019}) \bibinfo{pages}{753--767}.
\bibitem[{Oksanen(2006)}]{Oksanen2006}
\bibinfo{author}{J.~Oksanen}, \bibinfo{title}{Digital elevation model error in
  terrain analysis}, Ph.D. thesis, University of Helsinki,
  \bibinfo{address}{Helsinki}, \bibinfo{year}{2006}.
\bibitem[{Wechsler and Kroll(2006)}]{Wechsler2006}
\bibinfo{author}{S.~Wechsler}, \bibinfo{author}{C.~Kroll},
\newblock \bibinfo{title}{Quantifying {DEM} uncertainty and its effect on
  topographic parameters},
\newblock \bibinfo{journal}{Photogrammetric Engineering and Remote Sensing}
  \bibinfo{volume}{72} (\bibinfo{year}{2006}) \bibinfo{pages}{1081--1090}.
\bibitem[{Goovaerts(1997)}]{Goovaerts1997}
\bibinfo{author}{P.~Goovaerts}, \bibinfo{title}{Geostatistics for natural
  resources evaluation}, \bibinfo{publisher}{Oxford University Press},
  \bibinfo{year}{1997}.
\bibitem[{Temme et~al.(2009)Temme, Heuvelink, Schoorl, and
  Claessens}]{Temme2009}
\bibinfo{author}{A.~Temme}, \bibinfo{author}{G.~Heuvelink},
  \bibinfo{author}{J.~Schoorl}, \bibinfo{author}{L.~Claessens},
\newblock \bibinfo{title}{Chapter 5 geostatistical simulation and error
  propagation in geomorphometry},
\newblock in: \bibinfo{editor}{T.~Hengl}, \bibinfo{editor}{H.~Reuter} (Eds.),
  \bibinfo{booktitle}{Geomorphometry}, volume~\bibinfo{volume}{33},
  \bibinfo{publisher}{Elsevier}, \bibinfo{year}{2009}, pp.
  \bibinfo{pages}{121--140}.
\bibitem[{Kowalski et~al.(2018)Kowalski, Zhao, and Cai}]{Kowalski2018}
\bibinfo{author}{J.~Kowalski}, \bibinfo{author}{H.~Zhao},
  \bibinfo{author}{Y.~Cai},
\newblock \bibinfo{title}{Topographic uncertainty in avalanche simulations},
\newblock in: \bibinfo{booktitle}{International Snow Science Workshop
  Proceedings 2018}, pp. \bibinfo{pages}{690--695}.
\bibitem[{Zhao and Kowalski(2018)}]{Zhao2018}
\bibinfo{author}{H.~Zhao}, \bibinfo{author}{J.~Kowalski},
\newblock \bibinfo{title}{{DEM} uncertainty propagation in rapid flow-like
  landslide simulations},
\newblock in: \bibinfo{editor}{K.~Ho}, \bibinfo{editor}{A.~Leung},
  \bibinfo{editor}{J.~Kwan}, \bibinfo{editor}{R.~Koo}, \bibinfo{editor}{R.~Law}
  (Eds.), \bibinfo{booktitle}{Proceedings of the {Second} {JTC1} {Workshop} on
  Triggering and Propagation of Rapid Flow-like Landslides},
  \bibinfo{publisher}{The Hong Kong Geotechnical Society},
  \bibinfo{year}{2018}, pp. \bibinfo{pages}{191--194}.
\bibitem[{Remy et~al.(2009)Remy, Boucher, and Wu}]{Remy2009}
\bibinfo{author}{N.~Remy}, \bibinfo{author}{A.~Boucher},
  \bibinfo{author}{J.~Wu}, \bibinfo{title}{Applied Geostatistics with SGeMS},
  \bibinfo{publisher}{Cambridge University Press}, \bibinfo{year}{2009}.
\bibitem[{Lindsay(2018)}]{Lindsay2018}
\bibinfo{author}{J.~Lindsay}, \bibinfo{title}{WhiteboxTools user manual},
  \bibinfo{publisher}{Geomorphometry and Hydrogeomatics Research Group},
  \bibinfo{year}{2018}.
\bibitem[{{AECOM Asia Company Limited}(2012)}]{Report2012}
\bibinfo{author}{{AECOM Asia Company Limited}}, \bibinfo{title}{Detailed study
  of the 7 {June} 2008 landslides on the hillshade above {Yu Tung Road, Tung
  Chung}}, \bibinfo{type}{Technical Report}, \bibinfo{year}{2012}.
\bibitem[{DeBlasio and Elverhoi(2008)}]{DeBlasio2008}
\bibinfo{author}{F.~DeBlasio}, \bibinfo{author}{A.~Elverhoi},
\newblock \bibinfo{title}{A model for frictional melt production beneath large
  rock avalanches},
\newblock \bibinfo{journal}{Journal of Geophysical Research: Earth Surface}
  \bibinfo{volume}{113} (\bibinfo{year}{2008}).

\end{thebibliography}






\end{document}